\documentclass[aps,prd,preprint,preprintnumbers,unsortedaddress,superscriptaddress,showpacs,nofootinbib]{revtex4-1}
\usepackage{graphicx}
\usepackage{graphicx,color}
\usepackage{amsmath}
\usepackage{relsize}
\usepackage{slashed}
\usepackage{color}
\usepackage{ulem}
\def\lag{\langle}
\def\rag{\rangle}
\def\si{\sigma}
\def\ss{\lag\bar{s}s\rag}
\def\mixs{\lag\bar{s}g\si.Gs\rag}

\begin{document}

\title{Understanding close-lying exotic charmonia states within QCD sum rules}
\author{A.~Mart\'inez~Torres}
\affiliation{
Instituto de F\'isica, Universidade de S\~ao Paulo, C.P. 66318, 05389-970 S\~ao 
Paulo, SP, Brazil.
}
\author{K.~P.~Khemchandani}
 \affiliation{Faculdade de Tecnologia, Universidade do Estado do Rio de Janeiro,
Rod. Presidente Dutra Km 298, P\'olo Industrial, 27537-000, Resende, RJ, Brazil.
}
 \affiliation{Departamento de Ci\^encias Exatas e da Terra, Universidade Federal de S\~ao Paulo, Campus Diadema, Rua Prof. Artur Riedel, 275, Jd. Eldorado, 09972-270, Diadema, SP, Brazil.
}
\author{J. M. Dias}
 \affiliation{
Instituto de F\'isica, Universidade de S\~ao Paulo, C.P. 66318, 05389-970 S\~ao 
Paulo, SP, Brazil.
}
\author{ F.~S.~Navarra}
 \affiliation{
Instituto de F\'isica, Universidade de S\~ao Paulo, C.P. 66318, 05389-970 S\~ao 
Paulo, SP, Brazil.
}
\author{ M.~Nielsen}
 \affiliation{
Instituto de F\'isica, Universidade de S\~ao Paulo, C.P. 66318, 05389-970 S\~ao 
Paulo, SP, Brazil.
}

\preprint{}

\date{\today}

\begin{abstract}
Motivated by the experimental findings of some new exotic states decaying into channels like $J/\psi \phi$, we investigate the formation of resonances/bound states in the $D^*_s\bar D^*_s$ system using QCD sum rules. To do this
we start with a current of the type vector times vector and use spin projectors to separate the spin 0, 1 and 2 contributions to the
correlation function. We find three states with isospin 0, nearly spin degenerate, with a mass around 4.1 GeV. We have also investigated the decay of these states  to $J/\psi \phi$ and provide the corresponding partial widths. Such information should be useful for experimental studies in future.
\end{abstract}

\pacs{14.40.Rt,12.40.Yx, 13.75.Lb}

\maketitle
 
\section{Introduction}
In the last 13 years, a new flora of charmonium and botomonium like states have emerged from the $B$-factories: they are the so called $X$, $Y$ and $Z$ states. Having $X(3872)$ as the precursor, many more states came after it, as, for example, $X(4160)$, $X(4350)$, $Y(3940)$, $Y(4260)$, $Y(4660)$, $Z^+_c(3900)$, $Z^+_c(4020)$, $Z^+(4430)$. Some of these states are experimentally well established, as the case of $X(3872)$, and others not.  All these states decay to a set of open charm-anticharm mesons or a light and a hidden charm meson. All of them share the common feature of being good candidates for exotic mesons, with properties which do not match the conventional quark-antiquark structure.  Among all, the case of $Z$ states is even clearer since, in addition to bearing the decay property mentioned above,  they are electrically charged, which forces an internal structure of at least four quarks. This and other unconventional properties have originated the peculiar $XYZ$ nomenclature. 

These states are being widely studied and there are several experimental and theoretical reviews~\cite{godfrey,yuan,liu,nielsen,olsen,hosaka,Nielsen:2014mva}. The theoretical models employed to explain such states cover several color singlet configurations allowed by Quantum Chromodynamics (QCD) and are based on quark models~\cite{eef1,vijande,brodsky, braaten,esposito,chen,eef2}, effective field theories~\cite{ma,mko1,mko2,daniel,branz,ortega,wang,carlos,mknno1,liang,guo} and QCD sum rules calculations~\cite{albuquerque1,albuquerque2,Zhigangwang,kleiv,mknno2,mknn2,mo}.

In this manuscript we are going to study the possibility of forming states as a consequence of the dynamics involved in the $D^*_s\bar D^*_s$ system near its threshold, that is, around 4224 MeV. There exist two known $X$ states in this region: $X(4140)$ and $X(4160)$. In spite of the proximity of their masses, it seemed, until recently, that the two states are different since their widths were different. However, with a recent determination of the width of $X(4140)$, the two widths have now become compatible within the error bars, as we discuss below. 

$X(4160)$ was found by the Belle Collaboration~\cite{belle} in the $D^*\bar D^*$ invariant mass of the process $e^+e^-\to J/\psi D^*\bar D^*$. The mass and width obtained for this state are $4156\pm 29$ MeV and $139^{+ 113}_{-65}$ MeV, respectively.  Its quantum numbers (spin, isospin, parity) are not yet known. The decay of $X(4160)$ to $D^*\bar D^*$, in a $D_s^*\bar D_s^*$ bound state picture, could be understood in terms of triangular loop diagrams involving a strange meson.  Thus, $X(4160)$ could be a possible $D^*_s\bar D^*_s$ moleculelike state. In fact, several coupled channel studies of open charm meson systems  find a dynamically generated state and relate it to $X(4160)$~\cite{branz,liang,raquel}. Alternate interpretation of $X(4160)$ have also been presented,  in terms of different excited states in the $c\bar c$ spectra~\cite{li,yang,cao,zhu}, for instance, $\chi_0(3^3 P_0)$ ($J^{PC}=0^{++}$), $\chi_1(3^3 P_1)$ ($1^{++}$) or $\eta_{c2}(2^2D_2)$ ($2^{-+}$).  

The other state, $X(4140)$, was initially found by the CDF, CMS, and D0 Collaborations in the $J/\psi\phi$ invariant mass of the reactions $B^+\to J/\psi \phi K^+$, $p \bar p  \to J/\psi \phi +$ anything, with a mass around $4143$ 
 MeV and width $\sim 15^{+6}_{-5}$ 
MeV~\cite{cdf,cms,d0}. This state has recently been confirmed by the LHCb Collaboration too \cite{lhcb}. Although the mass value of $X(4140)$ determined in Ref.~\cite{lhcb} is in agreement with those found by the other Collaborations, the width obtained in Ref.~\cite{lhcb} is substantially larger: $\Gamma = 83 \pm 21^{+21}_{-14}$ MeV. Further, a Breit-Wigner fit analysis of the $J/\psi \phi$ mass spectrum made in Ref.~\cite{lhcb} seems to indicate the most preferred spin-parity of this state to be $1^{++}$.  However, the same study does not exclude an influence of the $D_s^{\pm} D_s^{*\mp}$ cusp effect in this energy region.  Actually some other works suggest that the  peak related to $X(4140)$ in the experimental data is not a real state but a kinematic effect~\cite{eef3,liuoka}. It is also worth mentioning that several other model calculations (quark models, effective field theories, QCD sum rules exploring different color singlet configurations allowed by QCD such as an excited $c\bar c$ state, $c\bar cs\bar s$ tetraquarks, $D^*_s\bar D^*_s$ molecule, etc.~\cite{gonzalez,liuzhu,lebed,albuquerque1,HidalgoDuque:2012pq}) predict the existence of a narrow state with mass around 4140 MeV with different  quantum numbers:  $J^{PC}=0^{++}$ or $1^{++}$ or $2^{++}$. To summarize: at present it is not clear if $X(4140)$ and $X(4160)$ are distinct states or not, and if the $X(4140)$ signal found in Ref.~\cite{lhcb} is affected by a nearby cusp or not. 

With the aim of contributing to this unconcluded  discussion, we perform a study of the $\bar D^*_s D^*_s$ system using QCD sum rules to explore the possible existence of the above mentioned states and their properties.  To proceed with our study we use the most straightforward current
for such a system, which is of the type vector current times vector current, and then use spin projectors to separate the several spin contributions to the two-point correlation function, as done in Refs.~\cite{mknno2,mknn2}. In this way the mass of the states is determined. Considering a three point correlation function and the spin projectors, a calculation of the decay width of the states found in our work, to the $J/\psi \phi$ channel, is performed. Our results can be useful for future experimental investigations.

\section{Determining the masses}
We start the calculations by writing the  simplest interpolating current  which can couple to the $D^{*-}_s D^{*+}_s$ system (which is directly in isospin $I=0$)
\begin{equation}\label{j}
j_{\mu \nu} (x) = \left[ \bar{c}_a(x) \gamma_\mu s_a(x)\right]\left[\bar{s}_b(x) \gamma_\nu c_b(x)\right],
\end{equation}
with $a,b$ denoting color indices. Using this current we determine the two-point correlation function 
\begin{equation}
\Pi_{\mu \nu \alpha \beta} (q^2) = i \int  d^4x e^{iqx} \langle 0 \mid T \left[ j_{\mu \nu} (x) j^\dagger_{\alpha \beta} (0) \right] \mid 0 \rangle \label{Pi},
\end{equation}
where $q$ is the four momentum at which the correlation function is evaluated and $T[\dots]$ denotes $T$-ordered product. Following Refs.~\cite{mknno2,mknn2}, we define the projectors
\begin{align} 
\mathcal{P}^{(0)}&=\frac{1}{3}\Delta^{\mu\nu}\Delta^{\alpha\beta},\nonumber\\
\mathcal{P}^{(1)}&=\frac{1}{2}\left(\Delta^{\mu\alpha}\Delta^{\nu\beta}-\Delta^{\mu\beta}\Delta^{\nu\alpha}\right),\label{proj}\\
\mathcal{P}^{(2)}&=\frac{1}{2}\left(\Delta^{\mu\alpha}\Delta^{\nu\beta}+\Delta^{\mu\beta}\Delta^{\nu\alpha}\right)-\frac{1}{3}\Delta^{\mu\nu}\Delta^{\alpha\beta},\nonumber
\end{align}
with
\begin{align}
\Delta_{\mu\nu}&=-g_{\mu\nu}+\frac{q_\mu q_\nu}{q^2}\quad(g^{\mu\nu}~\text{is the metric tensor}),\nonumber
\end{align}
which separate the spin 0, 1 and 2 contributions of the correlation function when the two hadrons forming the system interact in $S$-wave (angular momentum $L=0$). Thus, the projectors written in Eq.~(\ref{proj}) separate the $J^P=0^+, 1^+, 2^+$ contributions
 to the correlation function. If a state of isospin $I$, and spin $S$ is formed as a consequence of the interaction between a $\bar D^*_s$ and a  $D^*_s$ meson in angular momentum $L$, the parity $P$, C-parity and G-parity associated with the state found can be determined through $P=(-1)^L$, $C=(-1)^{L+S}$, and $G=(-1)^{L+S+I}$. In our case, since $I=0$ and $L=0$, the possible states must have positive parity and $C=G=(-1)^S$. This means states with quantum numbers $I^G (J^{PC})=0^+ (0^{++}),\,0^- (1^{+-}),\,0^+ (2^{++})$.

Once the correlation function is projected on spin, we need to rely on the dual nature of the correlation function from Eq.~(\ref{Pi}) to extract information from it: at short distances, the expression in Eq.~(\ref{Pi}) can be interpreted as quark-antiquark fluctuations, which is usually referred to as the QCD side, while at large distances it can be related to hadrons, which is  referred to as the phenomenological side. This is the key idea of the sum rule method, in which the correlation function is determined with both prescriptions and there should exist a range of $q^2$ where both results can be equated~\cite{svz,reviews1,reviews2}. 

When calculating the correlation function from the QCD side we have to deal with quark propagators and their expressions can be written using the operator product expansion (OPE),  with  the coefficients of the series calculated  perturbatively \cite{svz,reviews1,reviews2}. In practice, one calculates the spectral density which is associated with the correlation function through the dispersion relation
\begin{align}
 \Pi_{\textrm{OPE}}(q^2)=\int\limits_{s_{min}}^\infty ds \,\, \frac{\rho_{\textrm{OPE}}(s)}{s-q^2} + {\rm subtraction\,\, terms}.\label{corrope}
\end{align}

In the present case, we obtain the spectral density corresponding to the spin-projected correlation function by going in the OPE series up to terms with dimension eight 
\begin{align}
\rho^{S}_{\textrm{OPE}}&=\rho^{S}_{\textrm{pert}}+\rho^{S}_{m_s}+\rho^{S}_{\langle\bar s s\rangle}+\rho^{S}_{m_s \langle\bar s s\rangle}+\rho^{S}_{\langle g^2 G^2\rangle}+\rho^{S}_{\langle\bar s g\sigma G s\rangle}+\rho^{S}_{m_s\langle\bar s g\sigma G s\rangle}+\rho^{S}_{{\langle\bar s s\rangle}^2}\nonumber\\&~~+\rho^{S}_{m_s {\langle\bar s s\rangle}^2}+\rho^{S}_{\langle g^3 G^3\rangle} + \rho^{S}_{\langle\bar s s\rangle \langle\bar s g\sigma G s\rangle}+\rho^{S}_{m_s \langle\bar s s\rangle \langle\bar s g\sigma G s\rangle}.\label{rho}
\end{align}
The spin-projected OPE results are given in the appendix A of the paper. The numerical values used for the quark masses and condensates are given in Table~\ref{cond}~\cite{nielsen,reviews3}.
\begin{table}
\caption{Input quark masses and condensates for the OPE calculation of the correlation function.}\label{cond}
\begin{tabular}{cc}
\hline
Parameters&Values\\
\hline
$m_s$& $(0.13\pm0.03)$ GeV\\
$m_c$& $(1.23\pm 0.05 )$ GeV\\
$\langle q\bar q\rangle$&$-(0.23\pm0.03)^3$ GeV${}^3$\\
$\langle s\bar s\rangle$&$(0.8\pm0.2)\langle q\bar q\rangle$ \\
$\langle \bar s g \sigma\cdot G s\rangle$&0.8$\langle s\bar s\rangle$ GeV${}^2$\\
$\langle g^2 G^2\rangle$&$(0.88\pm0.25)$ GeV${}^4$\\
$\langle g^3 G^3\rangle$&$(0.58\pm0.18)$ GeV${}^6$
\end{tabular}
\end{table}

To determine the correlation function from the  phenomenological side we adopt the standard procedure of considering that the spectral density can be written as the contribution of a ``sharp" state (the one we are looking for), and the rest of possible excited states with same quantum numbers (the continuum):
\begin{align}
\rho^{S}_{\textrm{phenom}}(s)={\lambda^2_{S}}\delta(s-m_{S}^2)+\rho^{S}_{\textrm{cont}}(s). \label{rhopheno}
\end{align}
In Eq.~(\ref{rhopheno}), $S$ denotes the spin, $s = q^2$,  $\lambda_{S}$ is the coupling of the current to the state we are interested in and $m_{S}$ denotes its mass.
The density related to the continuum of states is assumed to vanish below a certain value of $s$, $s_0$, which is called continuum threshold. Above $s_0$ one usually  considers the ansatz~\cite{svz,reviews1,reviews2}
\begin{align}
\rho^S_{\textrm{cont}}(s)=\rho^{S}_{\textrm{OPE}}(s)\Theta(s-s_0).\label{ansatz}
\end{align}
Using this parametrization of the spectral density, the correlation function from the phenomenological side can be written as 
\begin{align}
 \Pi^{S}_{\textrm{phenom}}(q^2)=\frac{\lambda^2_{S}}{m_{S}^2-q^2}+\int_{s_0}^\infty ds\,\frac{\rho^{S}_{\textrm{OPE}}(s)}{s-q^2}.\label{corrphen}
\end{align}
To improve the matching between the two ways of calculating the correlation function, a Borel transformation is applied to Eqs.~(\ref{corrope}) and (\ref{corrphen}).  In this way, the contribution of the continuum on the phenomenological side  and divergent contributions arising due to long range interactions on the OPE side are both suppressed. Equating the Borel transformed results, we get the following expression for the mass 
\begin{align}
m_{S}^2=\frac{\int_{4m^2_c}^{s_0}ds\,s \rho^{S}_{\textrm{OPE}}(s)e^{-s/M^2_B}}{\int_{4m^2_c}^{s_0}ds\,\rho^{S}_{\textrm{OPE}}(s)e^{-s/M^2_B}},\label{mass}
\end{align}
and for  the coupling $\lambda_{S}$ 
\begin{align}
\lambda^2_{S}=\frac{\int_{4 m^2_c}^{s_0}ds\,\rho^{S}_{\textrm{OPE}}(s)e^{-s/M^2_B}}{e^{-m_{S}^2/M^2_B}},\label{lambda}
\end{align}
where $M_B$ is the Borel mass. The values of $s_0$ and $M_B$ are parameters of the theory and they are chosen to obtain a converging OPE series and a dominance of the ``sharp" state or pole term in the spectral density.

\begin{figure}[h!]
\includegraphics[width=0.43\textwidth]{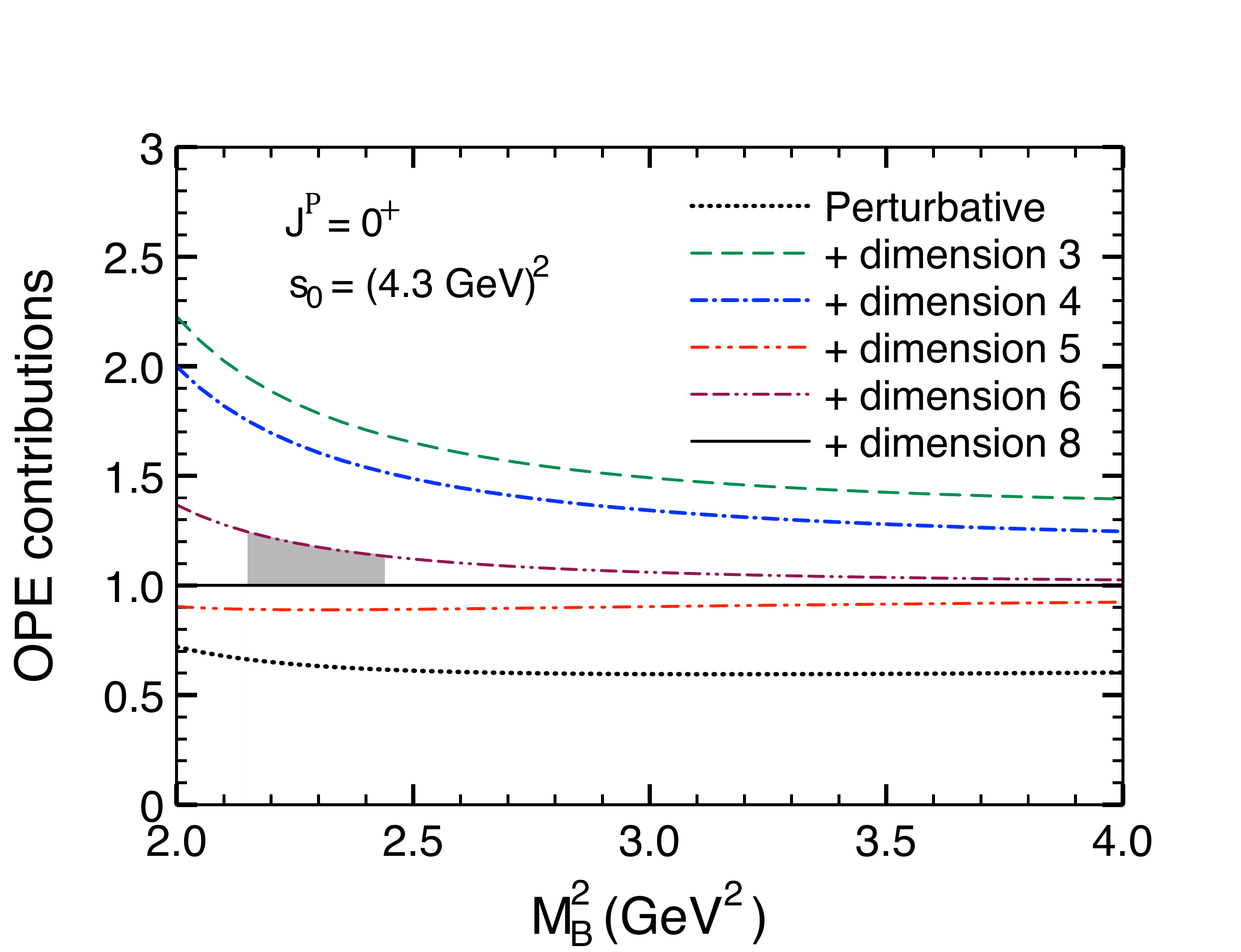}
\includegraphics[width=0.43\textwidth]{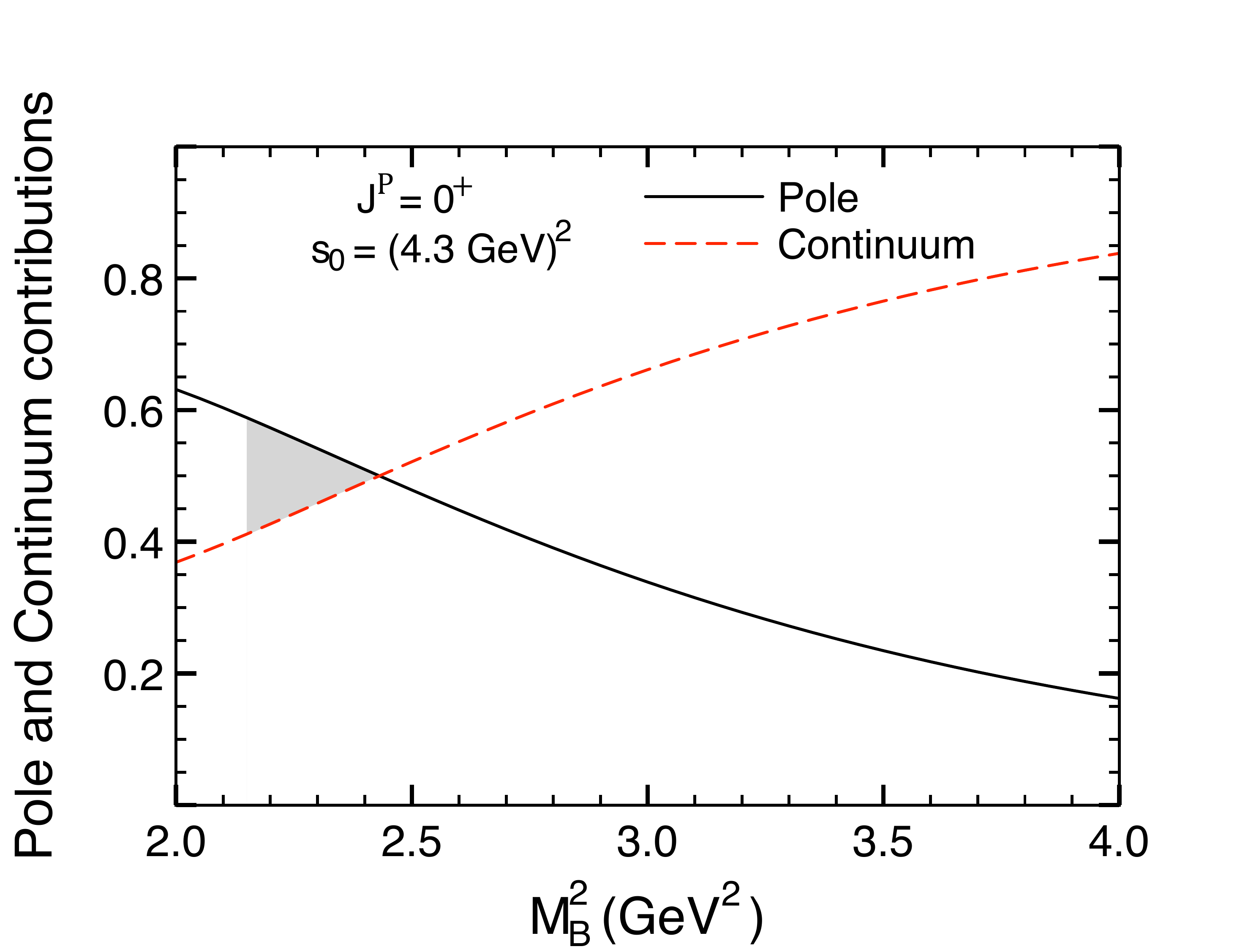}
\includegraphics[width=0.43\textwidth]{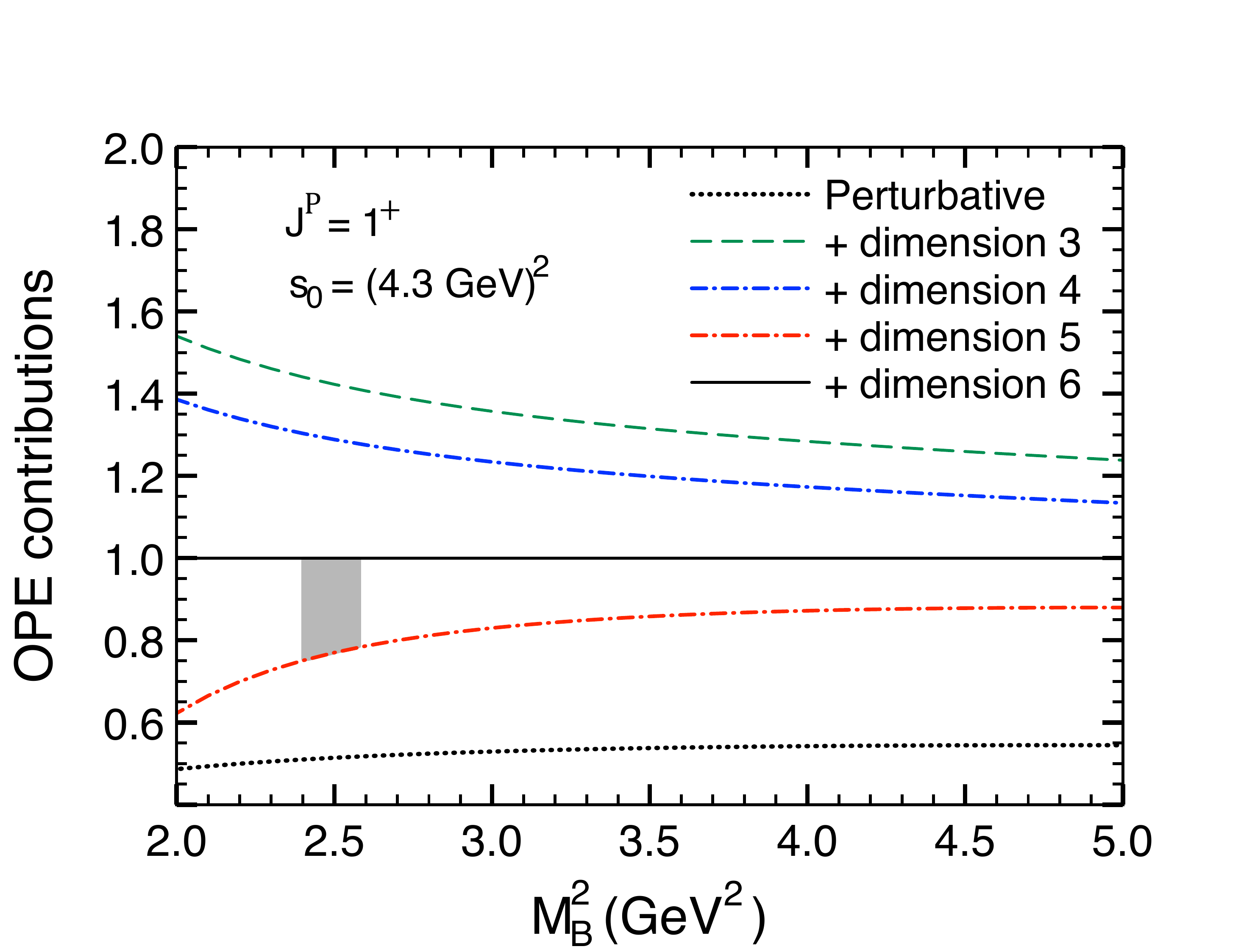}
\includegraphics[width=0.43\textwidth]{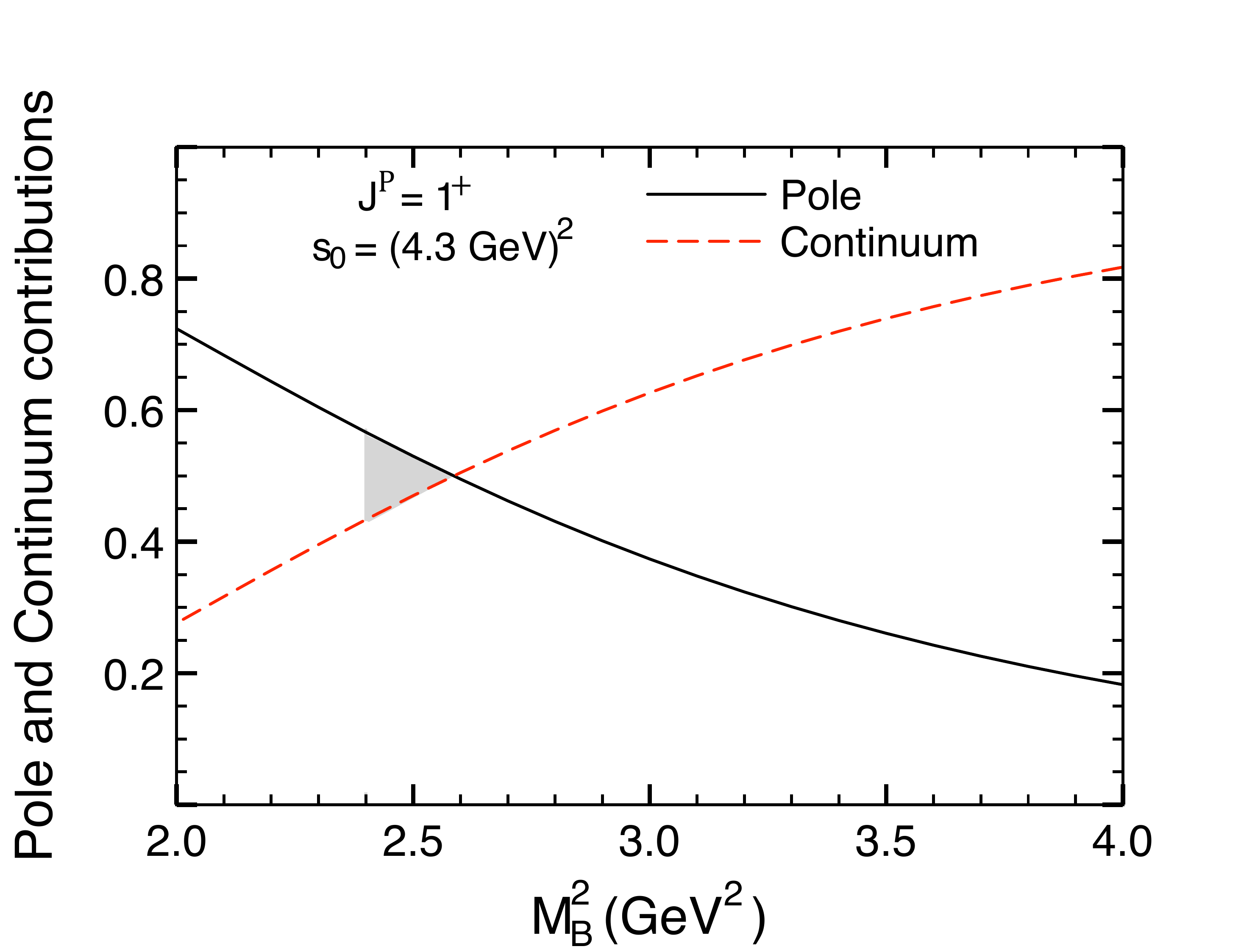}
\includegraphics[width=0.43\textwidth]{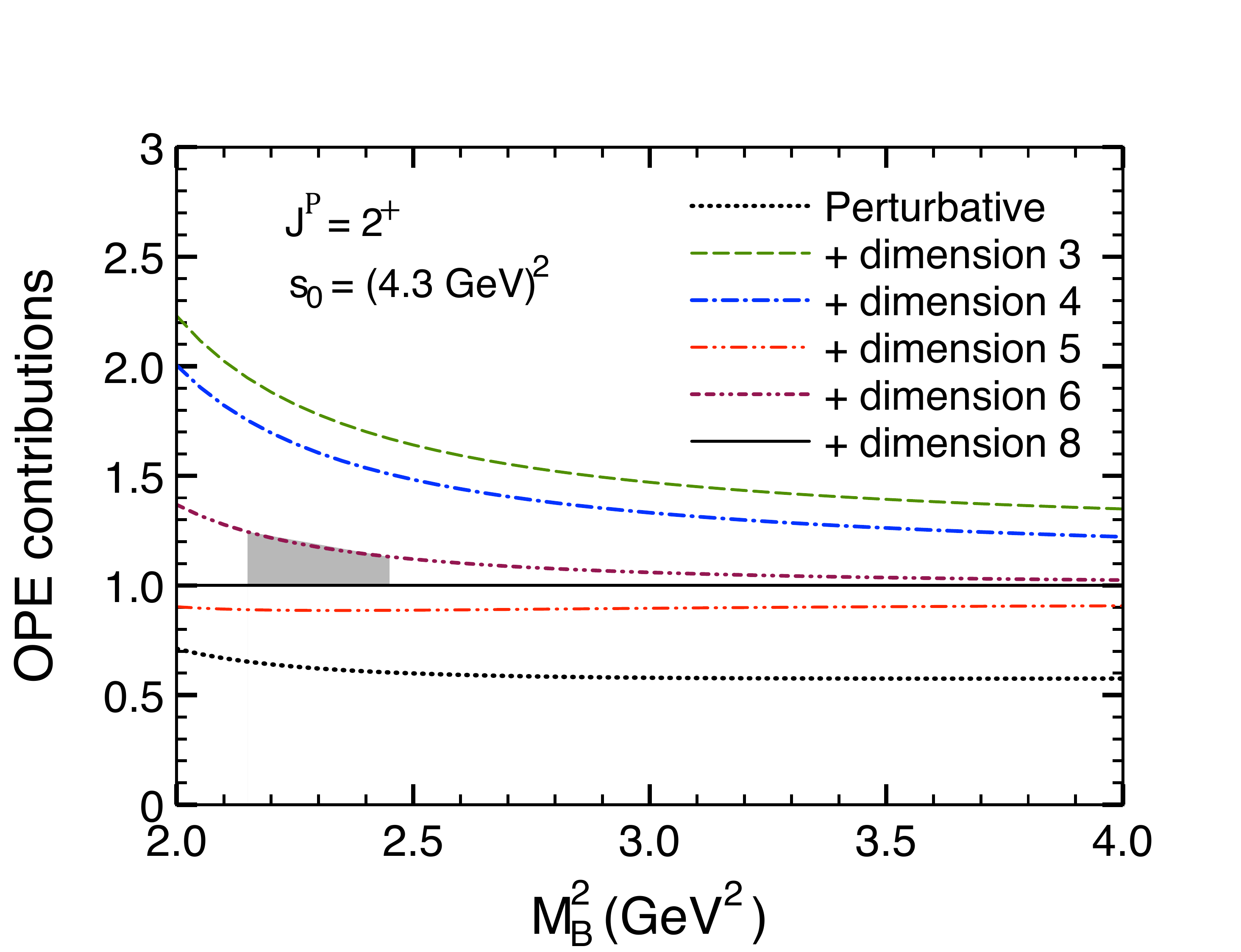}
\includegraphics[width=0.43\textwidth]{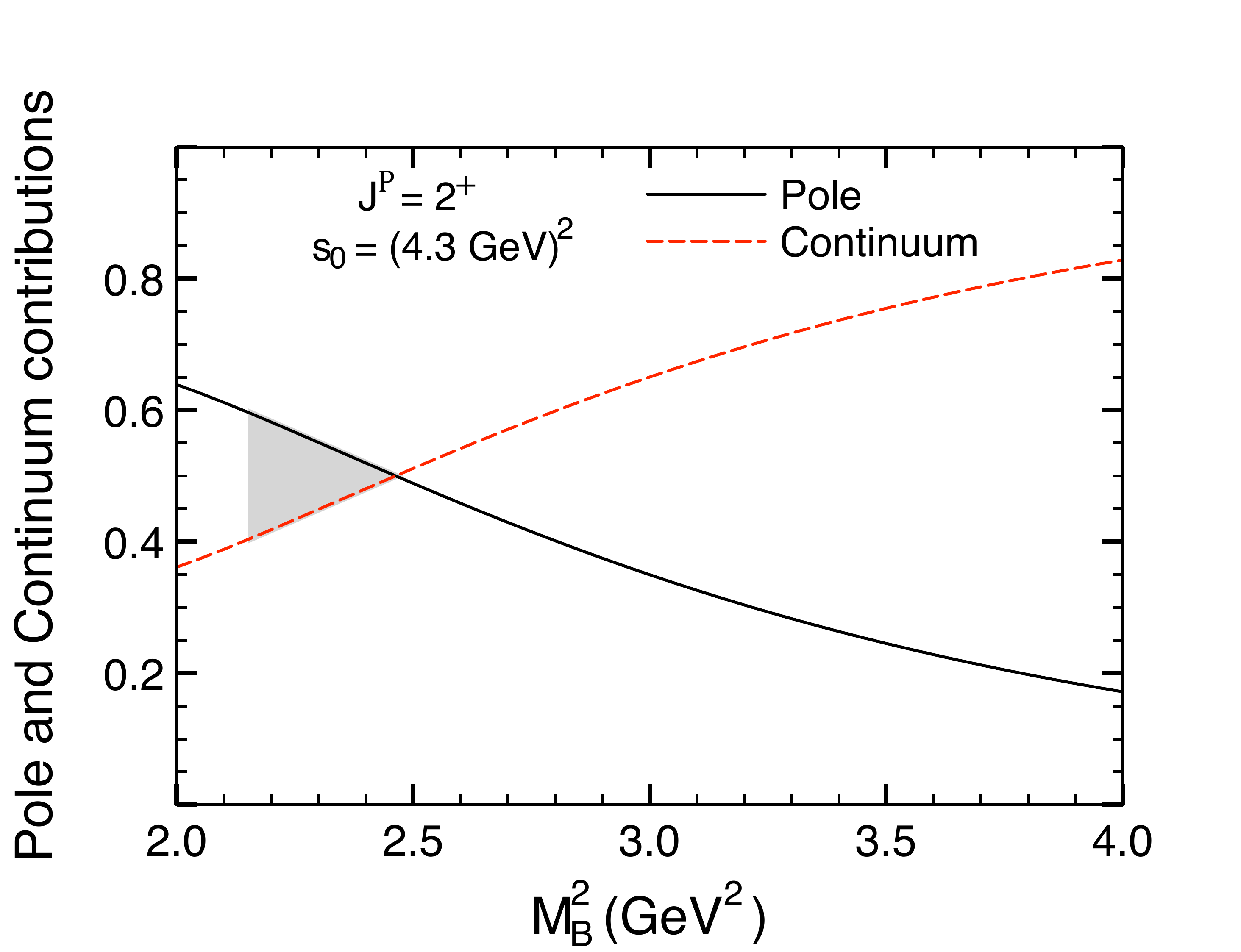}
\caption{(Left panels) Convergence of the OPE series as a function of the Borel mass squared $M^2_B$ for different $J^{\rm P}$ values. Each line represents the relative contribution of the specified term to the correlation function.  (Right panels) Dominance of the pole term over the continuum as a function of the Borel mass squared $M^2_B$ for different $J^{\rm P}$ values. Each line represents the relative contribution of the specified term to the correlation function. The shaded areas represent the Borel windows within which the conditions of  the convergence of the OPE series as well as the dominance of the pole term are satisfied.}\label{opepole}
\end{figure}

In Fig.~\ref{opepole} we show, 
respectively, the results found for the convergence of the OPE series and those for the dominance of the pole term for the case of spin $S = 0$,~1~and~2 and $\sqrt{s_0}=4.3$ GeV as a function of the Borel mass squared $M^2_B$. To obtain the results shown in Fig.~\ref{opepole} we have considered the central values associated with the quark masses and condensates shown in Table~\ref{cond}. As can be seen in the left panels, for instance, in case of $J^{\rm P} = 0^+$, for Borel mass of $M^2_B \sim 2.15$ GeV${}^2$, the difference between the contribution of the terms of dimension 6 and those of dimension 8 to the OPE series is around 25\%. This difference reduces further for higher values of $M_B$. Thus, we can consider this value of the Borel mass as the lowest one, beyond which a good convergence of the OPE series is ensured. From the right panel of the figure, we can see that there is a dominance of the pole over the continuum contribution for Borel masses smaller than $M^2_B \sim 2.45$ GeV${}^2$. It is then in the window of squared Borel masses $\left[2.15,2.45\right]$ GeV${}^2$ where we can guarantee a good convergence of the OPE series as well as the dominance of the pole term and only there it is meaningful to determine a mass and coupling from the expressions of Eqs.~(\ref{mass}) and (\ref{lambda}).

\begin{figure}[h!]
\includegraphics[width=0.7\textwidth]{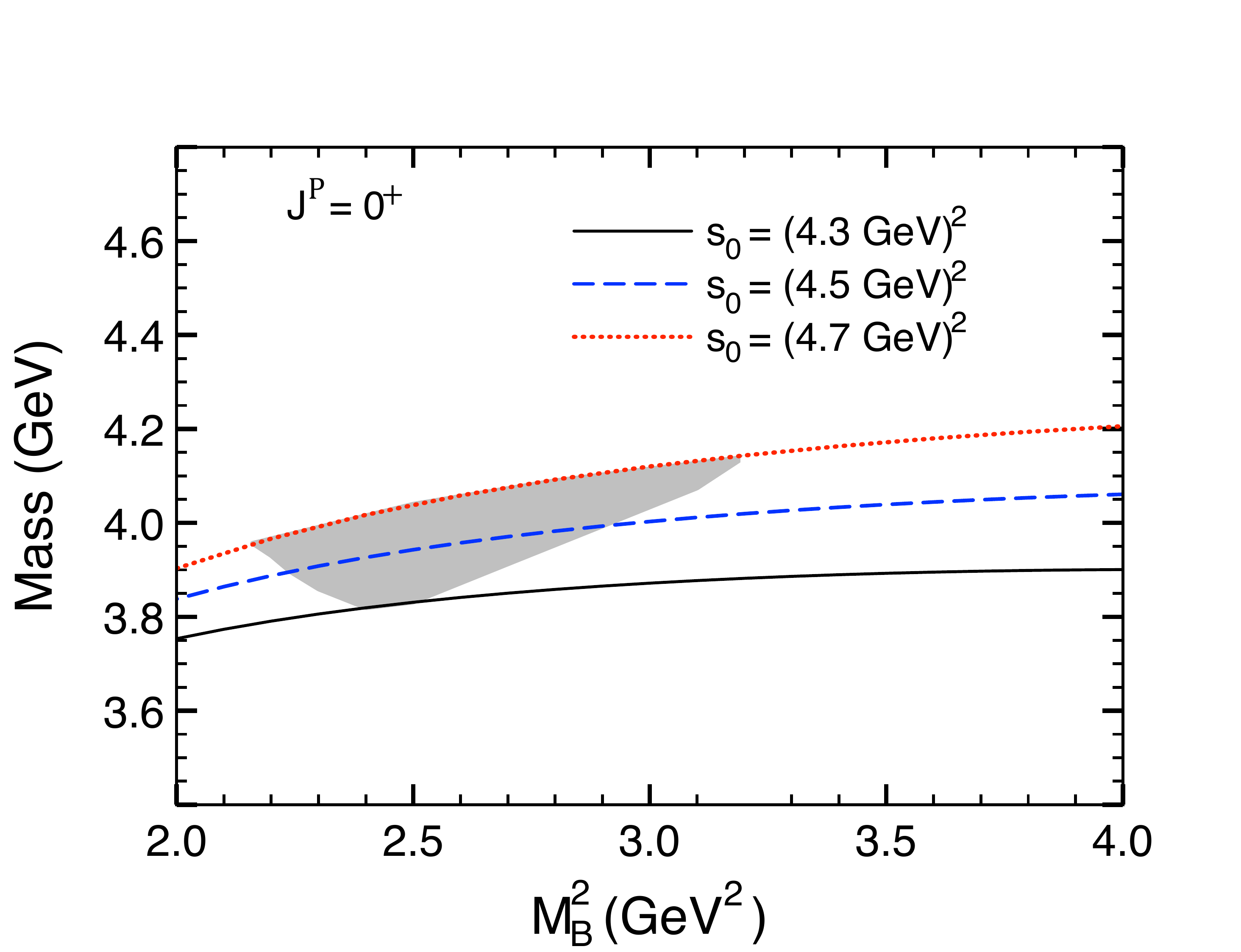}
\caption{Mass found for the spin 0 case for different values of the continuum threshold $\sqrt{s_0}$. The shaded area in the figure represents the allowed Borel region. }\label{mass0}
\end{figure}

In Fig.~\ref{mass0} we show the result for the mass found for the case of spin 0, for different values of $\sqrt{s_0}$, as a function of $M^2_B$. The shaded area shown in the figure represents the Borel window obtained in each case following the criteria mentioned above.  The values of the continuum threshold have been varied from the minimum value for which we have a valid Borel window, $\sqrt{s_0} = 4.3$ GeV, up to 4.7 GeV. In principle, one can take higher values for $\sqrt{s_0}$, but, as argued in Ref.~\cite{Zanetti:2016wjn}, this would imply a spectrum (on the phenomenological side) where the separation between the narrow ground state  and the continuum grows  larger and larger.  For example, a separation of $\sim$1 GeV between the ground state and the continuum certainly is a poor approximation. We, thus, consider 4.3 GeV $\le \sqrt{s_0} \le$ 4.7 GeV as a reasonable range of values. As can be seen from Fig.~\ref{mass0}, there is a good stability of the mass in the respective Borel windows and the mean value found is $M_0=(4.117 \pm 0.084)$~GeV. Similarly, for the coupling, we get $\lambda_0=(2.158 \pm 0.486) \times 10^{-2}$ GeV$^5$. The estimated errors here, and throughout the present manuscript,  correspond to two standard deviations from the mean value.

To estimate the errors, we must also consider other sources of uncertainties in the calculations, such as those associated to the quark masses and condensates listed in Table~\ref{cond}.  Taking all sources of errors into account, we get
\begin{align}
M_0=(4.114\pm 0.130)~\text{GeV},\quad \lambda_0=(2.215 \pm 0.564) \times 10^{-2}~\text{GeV}^5.\label{M0final}
\end{align}
Thus, the sum rule for the $\bar D^*_s D^*_s$ system in spin 0 suggests the formation of a state with quantum number $I^G (J^{PC})=0^+ (0^{++})$ and mass given by the value in Eq.~(\ref{M0final}).

\begin{figure}
\includegraphics[width=0.49\textwidth]{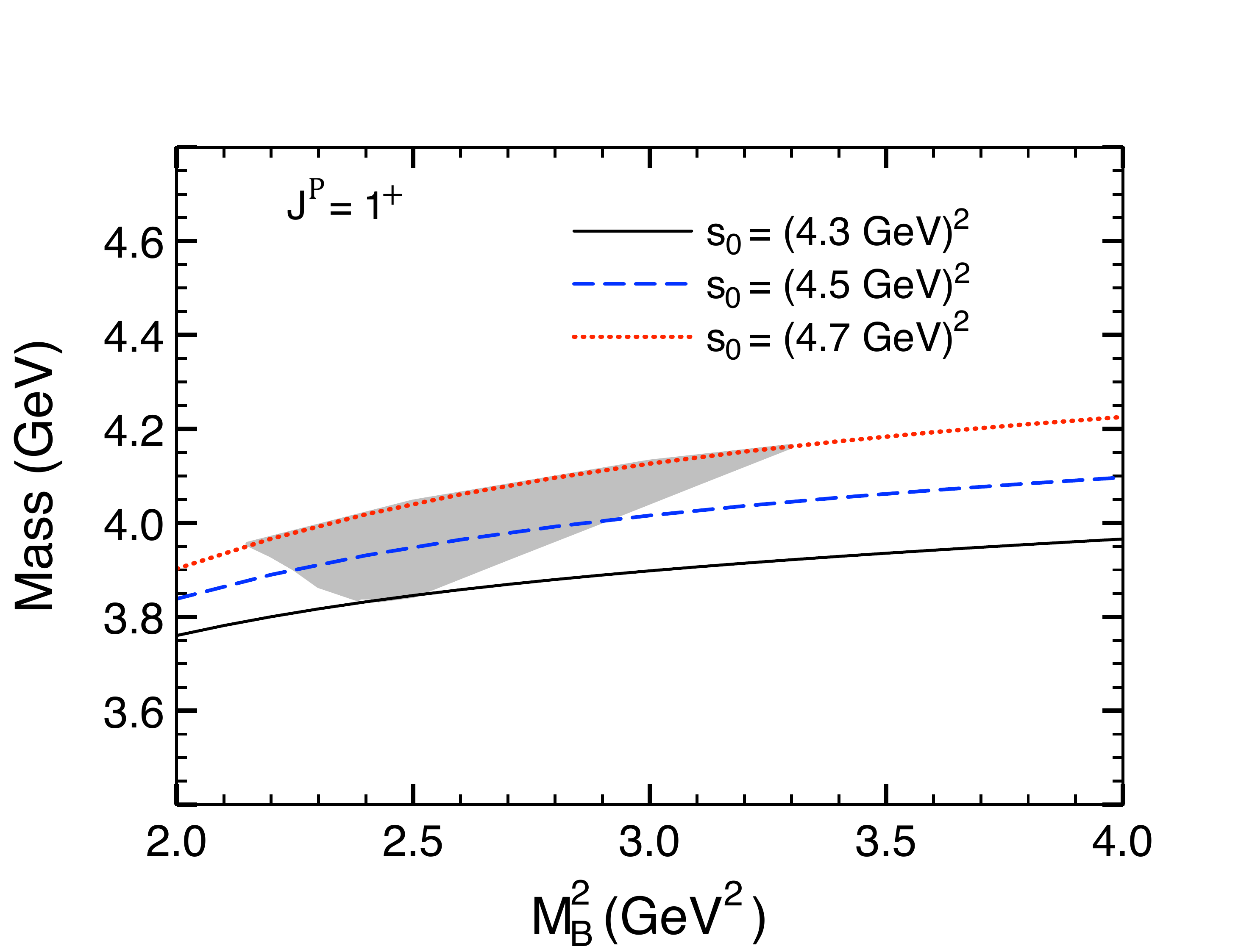}
\includegraphics[width=0.49\textwidth]{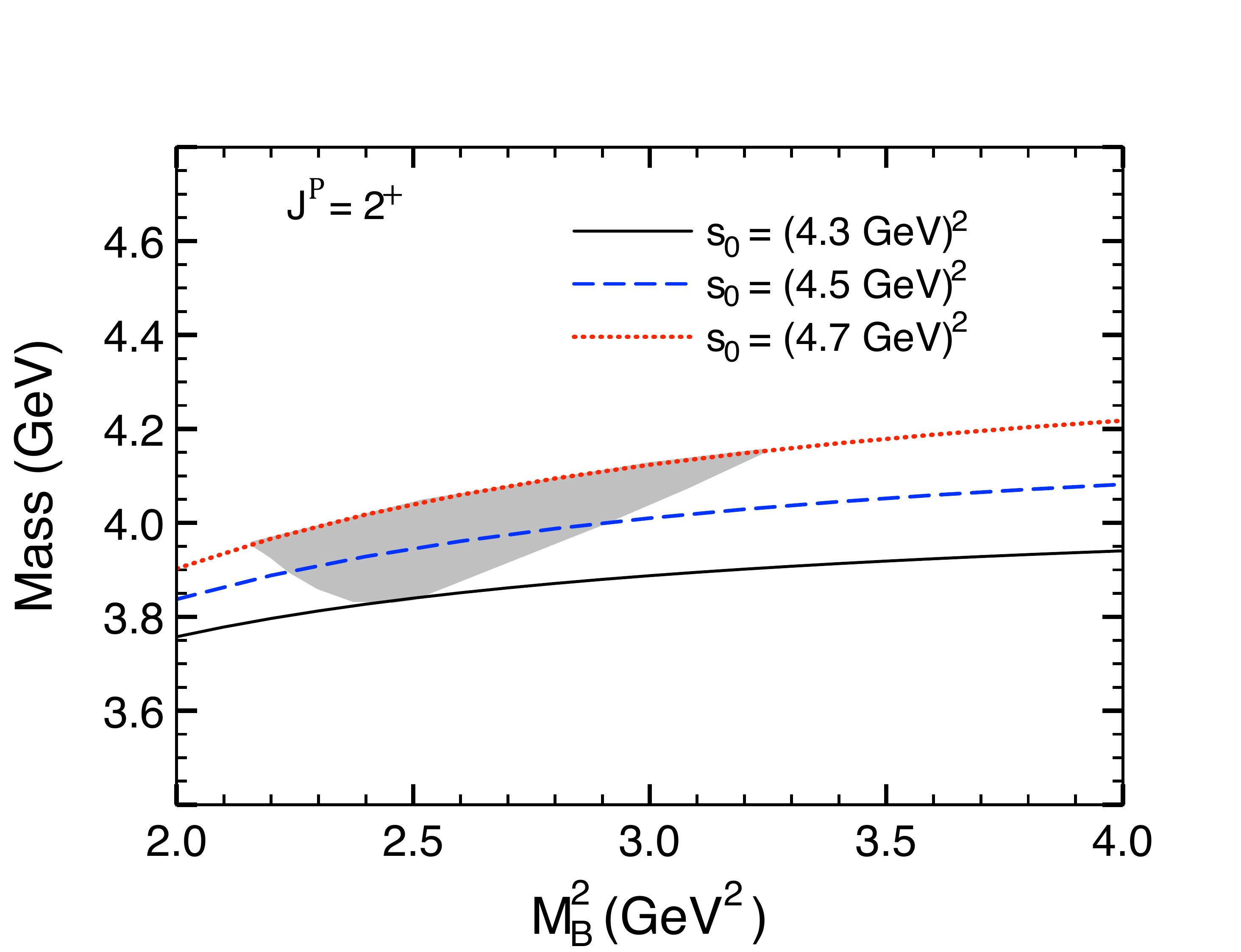}
\caption{Mass found for the spin 1 (left panel) and spin 2 (right panel) cases for different values of the continuum threshold $\sqrt{s_0}$. The shaded region in the figures has the same meaning as in Fig.~\ref{mass0}.}\label{mass12}
\end{figure}

Similarly, for the spin 1 and 2 cases, we also find a Borel window  (as shown in Fig.~\ref{opepole}) in which there is dominance of the pole term and good OPE convergence. The results for the central values of the parameters given in Table~\ref{cond}
are shown in Fig.~\ref{mass12} for several $\sqrt{s}_0$ as a function of the Borel mass $M_B$.

As can be seen from Fig.~\ref{mass12}, there is a stability for the masses in both spin $1$ and spin $2$ cases. Considering the uncertainties involved in the values of  the parameters of Table~\ref{cond}, we get the mean values for the masses of the states with spin $0$ and $2$ as
\begin{align}
M_1&=(4.120\pm0.127)~\text{GeV},\quad \lambda_1= (3.852 \pm 0.958) \times 10^{-2}~\text{GeV}^5\label{M1final}\\
M_2&=(4.117\pm 0.123)~\text{GeV},\quad \lambda_2=(4.963 \pm 1.243) \times 10^{-2}~\text{GeV}^5\label{M2final},
\end{align}
suggesting the presence of a $I^G (J^{PC})=0^- (1^{+-})$ and a $0^+(2^{++})$ states with masses given by the values in Eq.~(\ref{M1final}) and (\ref{M2final}), respectively.


It is worth mentioning that we have earlier studied  the $\bar D^* D^*$ system~\cite{mknn2} within the same formalism as presented here. Our study~\cite{mknn2} revealed the possibility of having spin $0$, $1$, and $2$ states around $3950 \pm 100$ MeV with isospin $0$ and $1$. Comparing the masses obtained in the study of the $\bar D^* D^*$ (isoscalar) current~\cite{mknn2} with  Eqs.~(\ref{M0final}), (\ref{M1final}), (\ref{M2final}), we see that, within the error bars, the two results are compatible. 

This result may appear surprising: the spin degeneracy found in each system for the mass of the state (within errors) could be understood on basis of the larger mass of the $c$ quark as compared to the one of the $u$, $d$, $s$ quarks, leading to an approximate heavy quark spin symmetry. Also, although the  central mass values of the states arising from a current having strange quarks (as is the case of the $\bar D^*_s D^*_s$ current) are larger when compared with those with light quarks ( the $\bar D^*D^*$ current), the difference  ($\sim 150$ MeV) is smaller than expected ($\sim 300$ MeV). From the point of view of a standard mass sum rule, as the one used here, the difference in studying the $\bar D^* D^*$ or the $\bar D^*_s D^*_s$ system arises from the fact that the latter one involves diagrams with strange quarks rather than the light ones. In the former case, the mass of the light quarks ($u$ and $d$) is taken to be negligible, thus, the corrections to the free propagator are related to condensates involving light quarks and gluons, while in the latter we have corrections  associated with strange quark condensates, gluon condensates and the mass of the quark $s$.  However, as can be seen from Table~\ref{cond}, the value for the light quark condensate is not very different from the one of the strange quark. Also, the $m_s$ corrections for the case of the $\bar D^*_s D^*_s$ system are not found to be large, as we have verified.  

However, this raises the following intriguing question: are the states found in the present work  the same as the ones found in $\bar D^* D^*$ system~\cite{mknn2} or do we have information on any new states? In principle, the bound states/resonances found in either case can couple to both $\bar D^* D^*$ and $\bar D^*_s D^*_s$ currents. One possible interpretation is that the states found with both currents are the same, which would lead to a conclusion that three states exist in the mass region 3.8-4.2 GeV, each with quantum numbers $0^{++}$, $1^{+-}$, $2^{++}$. These states can be interpreted as resonances/bound states coupling to both $\bar D^* D^*$  and $\bar D^*_s D^*_s$. 

Another possible interpretation is that the states found with  $\bar D^* D^*$ and $\bar D^*_s D^*_s$  currents are independent, and consequently our studies would imply the existence of 6 isoscalar states in the mass region $3.8-4.2$ GeV.  In fact, recent experimental findings of several isoscalar states in the $3.8-4.2$ GeV mass region ($X(3900)$, $X(3940)$, $X(4140)$, $X(4160)$ and $X(4230)$, etc.) may even indicate that the interpretation of  the states found in the $\bar D^* D^*$ and $\bar D^*_s D^*_s$ as being independent is compatible with Nature. However, as mentioned earlier, we must remember that the states found in  the $\bar D^* D^*$ system can couple to $\bar D^*_s D^*_s$, and vice versa. Thus, the states found in the spin projected currents of $\bar D^* D^*$ could also be found in a study of $\bar D^*_s D^*_s$ current and vice versa. In summary, the mass calculations alone do not seem to provide enough information to associate the states found in the present work with known states. 

 It is worth emphasizing at this point that the situation here is different when compared, for example, with the mass difference of $\rho$/$\omega-\phi$ mesons \cite{pascual}. In the case of the $\rho$/$\omega$ and $\phi$ mesons, we can write the current for the $\phi$-meson by replacing two light quarks in the $\rho$/$\omega$ current by two  strange quarks. Although the strange quarks are definitely heavier than the light quarks, when comparing with $D^*\bar D^*$ and $D_s^*\bar D_s^*$, we are dealing with two charm quarks in both systems and the fact that charm quarks are much heavier than both the strange and the light quarks makes that the presence of the charm quarks overshadows the mass difference between light and strange quarks. This can be compared to the consideration of heavy quark symmetry in the calculations based on effective field theories where  $D^*\bar D^*$ and $D_s^*\bar D_s^*$ are considered as coupled channels and, indeed, the results obtained in such calculations \cite{HidalgoDuque:2012pq} seem to be very similar to ours.

We must also mention that there exist other calculations which show that the heavy quark symmetry may not always hold and it may be important to consider mixing of interactions in the s- and  d- partial waves \cite{Bondar:2011ev,Voloshin:2011qa}, like in nucleon-nucleon interactions.   It might be useful to investigate the effect of s-d partial wave mixing in QCD-sum rules in future calculations (such effects have been considered in QCD sum rules, for example, in Ref.~\cite{Reinders:1984sr}).


 In view of the  current situation, though, we need more information than just the masses of the states. It should, thus, be useful to study the partial decay widths of  the states found in our work to those channels which can be studied experimentally. 
A comparison between the results and data might then be helpful in identifying which state(s) can be interpreted as a resonance/bound state of the open charm vector mesons. Since the states found in the present work are of hidden charm and, hidden strange nature, it should be ideal to study their decay to a channel like $J/\psi~\phi$. The decay of the $D^* \bar D^*$ states to such a channel is OZI suppressed and expected to be small. Such analysis should be helpful in distinguishing the states coupling to $D^* \bar D^*$ and $D_s^* \bar D_s^*$.

\section{Determining the decay widths to $J/\psi~\phi$}

The starting point in the width calculation within the QCD sum rules approach is the determination of the 
coupling constant for the $XJ/\psi\phi$ vertex, where $X$ represents the spin 0 or spin 2 state (our $1^{+-}$ state cannot decay to $J/\psi \phi$). Once we get the coupling constant value, it is possible to estimate the decay width of our states to $ J/\psi\phi$.

The procedure begins with the determination of the three-point correlation function 
\begin{equation}\label{3point}
\Pi_{\mu \nu \alpha \beta} (p^2)= \int d^4x\, d^4y e^{i\,p^\prime \cdot x} e^{i\, q \cdot y} \Pi_{\mu \nu \alpha \beta}(x,y)\, ,
\end{equation}
where $p=p^{\prime}+q$ is the four momenta of the incoming particle, $p^\prime$ and $q$ are, respectively, four momenta related to $J/\psi$ and $\phi$, and $\Pi_{\mu \nu \alpha \beta}(x,y)$ is given by
\begin{equation}
\Pi_{\mu \nu \alpha \beta}(x,y) = \langle 0 | T\{j_{\mu}^{\psi}(x)\,j_{\nu}^{\phi}(y)\, j_{\alpha \beta}^{\dagger X}(0) \} | 0 \rangle \, .\label{pij}
\end{equation}
In Eq.~(\ref{pij}), $j_{\nu}^{\phi}$ and $j_{\mu}^{\psi}$ correspond to the interpolating currents for the $\phi$ and $J/\psi$ mesons and they are defined as
\begin{eqnarray}\label{j's}
j_{\mu}^{\psi} = \bar{c}_a(x)\gamma_{\mu} c_a(x)\, ,\nonumber\\
j_{\nu}^{\phi}=\bar{s}_b(y)\gamma_{\nu} s_b(y)\, ,
\end{eqnarray}
with $a$ and $b$ representing color indices. The interpolating current associated with the $X$ meson, $j_{\alpha \beta}^{X}$, can be found in Eq.~\eqref{j}.

As in the case of the correlation function for the mass, the quark-hadron duality is used to evaluate Eq.~(\ref{3point}). The phenomenological side is calculated considering that the interpolating currents present in Eq.~(\ref{pij}) create/annihilate the corresponding hadrons, while the OPE side is evaluated in terms of quark condensates, gluon condensates, etc. Once again, a Borel transformation is done to improve the matching between the two descriptions. 

We start our discussion with the determination of the decay width of the spin 0 state  found in our work.

To determine the phenomenological side of Eq.~(\ref{3point}), we start by inserting in it a complete set of intermediate states for the $X$, $J/\psi$ and $\phi$ mesons and use the following definitions~\cite{albuquerque2,mknno2}
\begin{eqnarray}
\langle 0|\, j_{\mu}^{\psi}\,|\psi(p^{\prime})\rangle= m_{\psi}f_{\psi}\epsilon_{\mu}(p^{\prime}),\nonumber\\
\langle 0|\, j_{\nu}^{\phi}\,|\phi(q)\rangle= m_{\phi}f_{\phi}\epsilon_{\nu}(q),\nonumber\\
\langle 0|\,j_{\alpha \beta}\,|X(p)\rangle =\lambda_0\Big(g_{\alpha\beta}+ \frac{p_{\alpha}p_\beta}{p^2}\Big).\label{defphen}
\end{eqnarray}
The coupling $\lambda_0$ in Eq.~(\ref{defphen}) is precisely the coupling constant found for the spin 0 state, Eq.~(\ref{M0final}).
Using the spin 0 projector of Eq.~(\ref{proj}) and the fact that now the four momentum at which the correlation  function of Eq.~(\ref{3point}) is evaluated corresponds to $p^2$, we get the following expression for the correlation function in spin 0
\begin{align}\label{ps1}
\Pi^{0}_\text{phenom}&=\frac{f_{\psi} f_{\phi} \lambda_0 m_{\psi} m_{\phi}\, g_{X\psi\phi}(q^2)}{p^2(p^{\prime 2} - m_{\psi}^2)(p^2 - m_X^2)
(q^2 - m_{\phi}^2)}\Big\{ (p\cdot p^{\prime})\,(p\cdot q)+2 p^2 (p^{\prime}\cdot q) \Big\}\,+\, .\, .\, .\, ,
\end{align}
with $m_{\psi}$, $m_{\phi}$ and $f_{\psi}$, $f_{\phi}$ being, respectively, the masses and decay constants of the $J/\psi$ and $\phi$ mesons. In Eq.~(\ref{ps1}), the dots represent the contribution from the excited states (which includes pole-continuum and continuum-continuum contributions) and $g_{X\psi\phi}(q^2)$ is a form factor for the vertex $XJ/\psi\phi$. This form factor is defined by the generalization of the  on-mass-shell matrix element, $\mathcal{M}$, for $X \to J\psi\phi$ to a case in which the $\phi$ meson is off-shell with four momenta $q$, with
\begin{align}
 \mathcal{M}_{XJ\psi\phi}&=g_{X\psi\phi}(q^2)\, \Big\{ [p^{\prime}\cdot \epsilon^*(q)] [q\cdot \epsilon^*(p^{\prime})]
- [p^{\prime}\cdot q][\epsilon^*(p^{\prime})\cdot \epsilon^*(q)] \Big\}\, ,
\end{align}
and it can be obtained from an effective Lagrangian describing the $XJ/\psi \phi$ vertex. Since in this case the hadron $X$ is a scalar, a Lagrangian for the $XJ\psi \phi$ vertex is given by~\cite{albuquerque2}
\begin{equation}
\mathcal{L}=\frac{1}{2}\,\mathcal{G}_{X\psi\phi}\,V_{\alpha\beta}\,\psi^{\alpha\beta} X\,,\label{x}
\end{equation}
where $V_{\alpha\beta}=\partial_{\alpha}\phi_{\beta}-\partial_{\beta}\phi_{\alpha}$ and $\psi^{\alpha\beta}=\partial^{\alpha}\psi^{\beta}-\partial^{\beta}\psi^{\alpha}$ are tensor fields associated with the $\phi$ and $J/\psi$ mesons, respectively,  $X$ in Eq.~(\ref{x}) represents the scalar field for the spin 0 state and $\mathcal{G}_{X\psi\phi}$ the coupling constant for the vertex, which is related to the form factor $g_{X\psi\phi}(q^2)$.

On the OPE side, we calculate Eq.~\eqref{3point} at leading order in $\alpha_s$ and consider condensates up to dimension 5. Then, using the spin 0 projector of Eq.~(\ref{proj}) we extract the spin 0 part of the correlation function. After this, a Borel transformation is performed in the OPE side as well as in the phenomenological side and a matching between the two results is done. As a result, we find the following relation \cite{ioffe,eidemuller}

\begin{eqnarray}\label{sumrule}
g_{X\psi\,\phi}(Q^2)\,A(M_B^2,Q^2)\,+\, B\,e^{-u_0/M_B^2}=\int\limits_{4m_c^2}^{s_0}\, ds \, e^{-s/M^2_B}\rho^{0}_\text{OPE}(s,Q^2).
\end{eqnarray}
To obtain Eq.~(\ref{sumrule}), since $X$ and $J/\psi$ are much heavier than $\phi$, we consider the approximation $p^2\simeq {p^\prime}^2$~\cite{Bracco:2011pg, albuquerque2}.   A brief discussion in order is here. The approximation $p^2\simeq {p^\prime}^2$ may lead to a misinterpretation that the uncertainty implied by such an approximation can be of the order of the  mass difference between $J/\psi$ and  $X$. It is important to recall that, although the approximation is motivated by the proximity of the mass values of $X$ and $J/\psi$ (as compared to $\phi$),  both $p$ and $p^\prime$ are off-shell variables in the three-point correlation function and, thus, the uncertainty in the results is not as large as the mass difference of $J/\psi$ and  $X$. The approximation   $p^2\simeq {p^\prime}^2$ is a very common one, used when calculating  three-point correlation functions where two of the particles involved have masses of similar order (see, for example, Ref.~\cite{ioffe}). An alternative procedure would consist of, for instance, applying a double Borel transformation to the correlation function (see the review~\cite{Bracco:2011pg}  for more detailed discussions), which would result in the introduction of two parameters in the form of Borel masses. In such a case, one would have to impose some kind of relation between the two Borel masses to avoid two free parameters.
There is no way, a priori, of knowing which method is better,  and to avoid such additional uncertainties, we choose to stick to the $p^2\simeq {p^\prime}^2$ approximation and have one Borel mass parameter.

Proceeding further, the Borel transformation is performed for the Euclidean variable $P^2=-p^2$, which introduces a Borel mass $M_B$ (as in case of the sum rule for the mass). In Eq.~(\ref{sumrule}), $Q^2$ corresponds to the Euclidean variable related to $q^2$ ($Q^2=-q^2$), $B$ represents contributions from pole-continuum and continuum-continuum transitions~\cite{eidemuller}, $u_0$ and $s_0$ are the continuum threshold parameters for $X$ and $J/\psi$, respectively, and
\begin{align}
\rho^{0}_\text{OPE}(s,Q^2)=\rho^{0}_\text{pert}(s,Q^2)+\rho^{0}_{\ss}(s,Q^2)+\rho^{0}_{\mixs}(s,Q^2)\label{jope}
\end{align}
is the spectral density found on the OPE side after doing the above mentioned approximations and the projection on spin 0. The density in Eq.~(\ref{jope}) considers perturbative, quark and mixed condensates contributions in the OPE side and the corresponding expressions can be found in the Appendix B. Next, $A(M_B^2,Q^2)$ in Eq.~(\ref{sumrule}) is given by
\begin{align}\label{phenS0}
A(M_B^2,Q^2)&=\frac{f_{\psi}f_{\phi}\lambda_0 m_{\phi} Q^2 e^{-m_{\psi}^2/M_B^2}e^{m_X^2/M_B^2}}{4m_{\psi} m_X^2(m_X^2-m_{\psi}^2)(m_{\phi}^2+Q^2)}\Big\{m^2_X e^{m_X^2/M_B^2} (Q^2-2m^2_{\psi})+ e^{m_{\psi}^2/M_B^2}\nonumber\\
&\times [Q^2 e^{m_X^2/M_B^2}(m^2_{\psi}-m^2_X)+m^2_{\psi}(2m^2_X-Q^2)] \Big\}\, .
\end{align}
Using Eq.~(\ref{sumrule}) and its derivative with respect $1/M^2_B$, we can eliminate the unknown $B$ and obtain in this way an expression for the form factor $g_{X\psi\,\phi}(Q^2)$ which can be calculated numerically.  The calculations are done for a range of $Q^2$  and $M^2_B$  and a test of the reliability of the results is made by analyzing the Borel stability of the form factor calculated using Eq.~(\ref{sumrule}) as a function of $Q^2$ and $M^2_B$.

\begin{figure}[h!]
\includegraphics[width=0.5\textwidth]{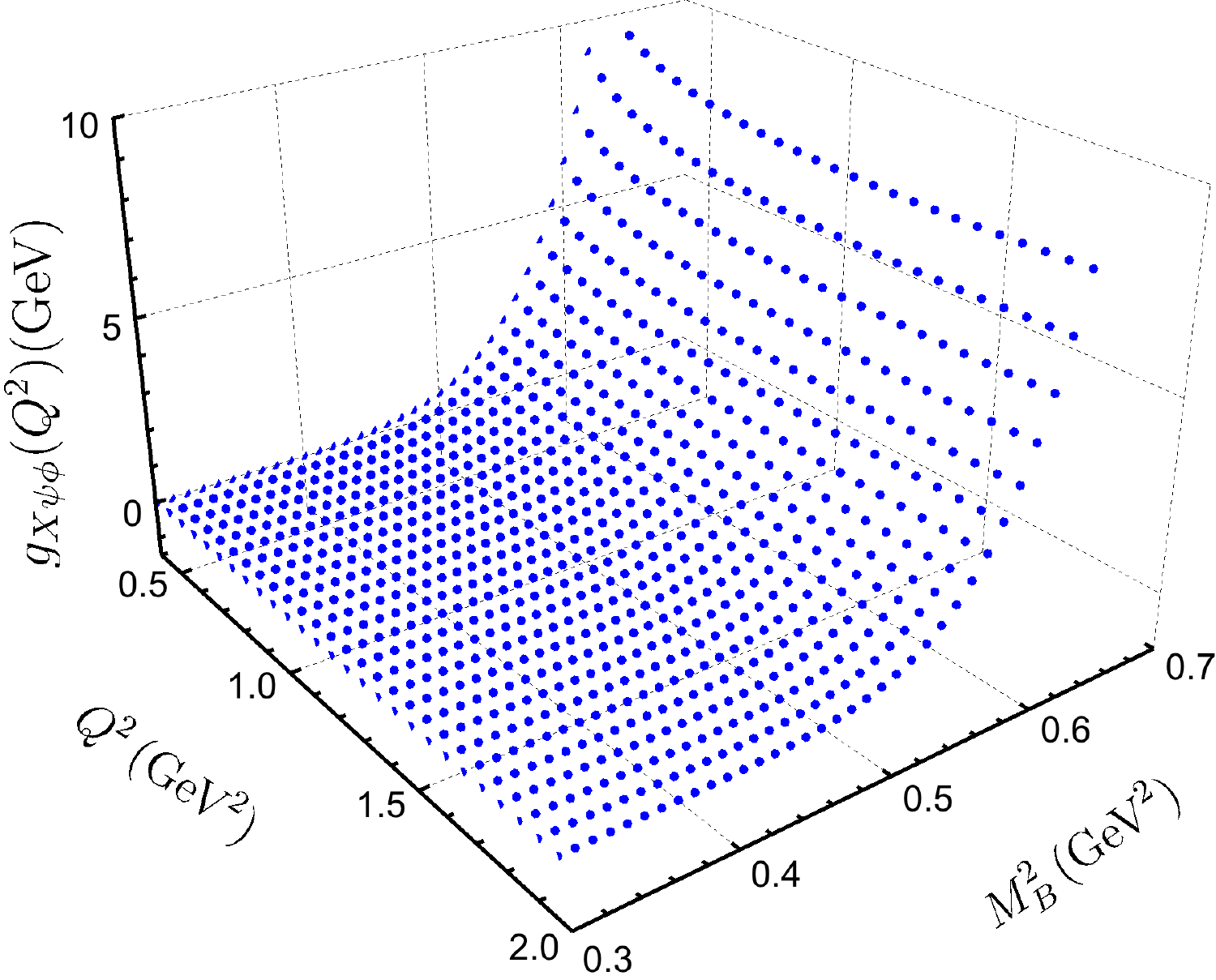}
\caption{Numerical solution of Eq.~\eqref{sumrule}, which provides the form factor 
$\mathcal{G}_{X\psi\,\phi}(Q^2)$, for the spin 0 case, as a function of $M^2_B$ as well as $Q^2$.}\label{g3DS0}
\end{figure}

In Fig.~\ref{g3DS0} 
we show such a stable region of the form factor $g_{X \psi \phi}$, which is associated with the vertex $X J/\psi \phi$ (with $X$ having spin 0). It can be seen that $g_{X \psi \phi}$ exhibits a plateau for values of $M^2_B$ between 0.3-0.5 GeV$^2$.

\begin{figure}[h!]
\includegraphics[width=0.5\textwidth]{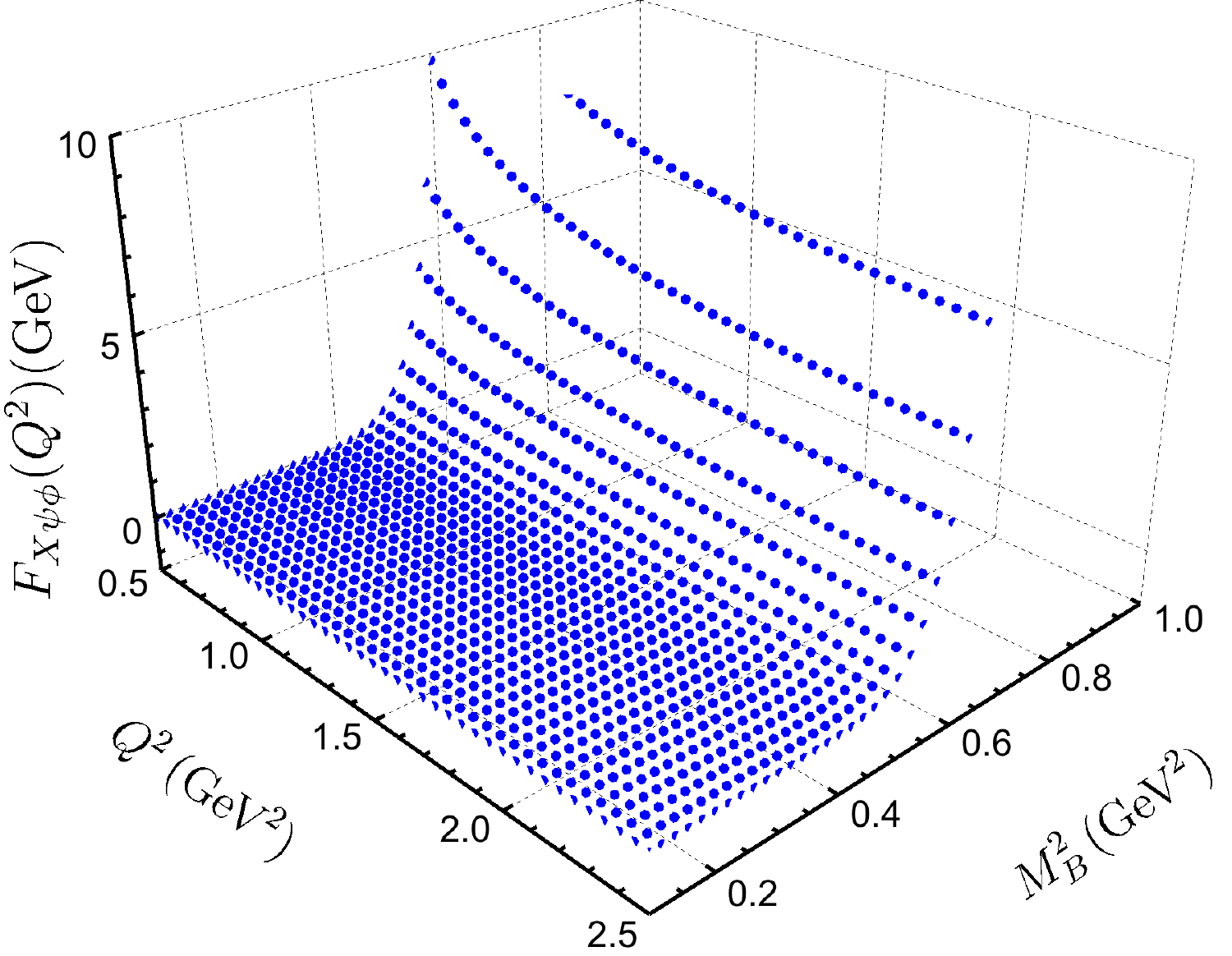}
\caption{Numerical solution of Eq.~\eqref{sumrule_s2}, which provides the form factor 
$\mathcal{F}_{X\psi\,\phi}(Q^2)$, for spin 2 case, as a function of $M^2_B$ as well as $Q^2$.}\label{g3DS2}
\end{figure}

Similarly, as shown in Fig.~\ref{g3DS2}, the form factor $F_{X \psi \phi}$, which is  related to the vertex $X J/\psi \phi$ (with $X$ being a spin 2 hadron), manifests a plateau in the interval 0.1 GeV$^2 \le M^2_B \le$ 0.6 GeV$^2$.

The finding of a plateau is an indication of the weak dependence of the sum rule with the Borel mass parameter. Thus, to calculate the coupling constants needed for the evaluation of the decay widths, we can fix the value of $M^2_B$ to one of the values in the plateau region and perform the extrapolation of the form factor from $Q^2$ in the space-like region (where the sum rule is evaluated) to $Q^2 = - m^2_\phi$ in the time-like region (where the coupling constant is determined)~\cite{Bracco:2011pg}.

\begin{figure}[h!]
\includegraphics[width=0.6\textwidth]{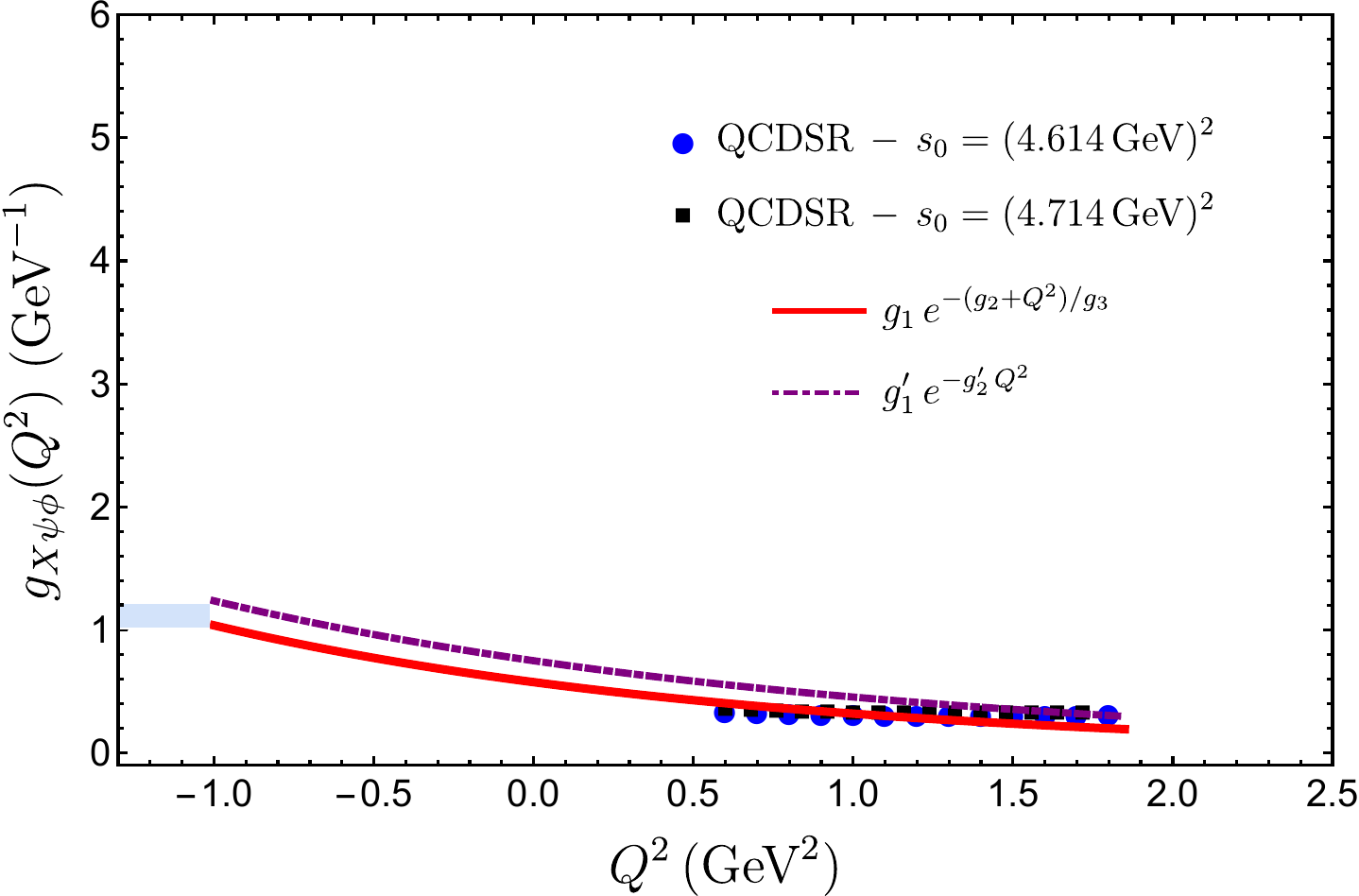}
\caption{The filled circles and boxes in the  $Q^2>0$ region represent the results of the QCD sum rule calculation of $g_{X\psi\phi}$ for the state with spin 0 as a function of  $Q^2$ for two different values of the continuum threshold, $s_0$. These results correspond to a value of  $M^2_B$ which belongs to the stable region shown in Fig.~\ref{g3DS0}. The solid and dashed lines correspond to the results obtained with the form factor given in Eqs.~\eqref{ffa}-\eqref{ffb}. The shaded region near the vertical axis represents the range of the values for the coupling constant $\mathcal{G}_{X\psi\phi}$ obtained by the extrapolation procedure explained in the text.}\label{gS0}
\end{figure}

We show such an extrapolation  for the form factor  $g_{X \psi \phi}$ in Fig.~\ref{gS0}, for a fixed value of $M^2_B$ which lies in the stable plateau region shown in Fig.~\ref{g3DS0}.  The filled circles and boxes correspond to the sum rule results for two different values of $s_0$, which give an idea of the uncertainties in the results related to the value of the continuum threshold. To determine the coupling constant, which corresponds to the form factor at the meson pole, we extrapolate the sum rule results, by fitting them with the following  form factors \cite{Bracco:2011pg}:
\begin{align}\label{ffa}
g_{X\psi\phi}(Q^2)=g_1\,e^{-\frac{(g_2+Q^2)}{g_3}}\, ,\\\label{ffb}
g_{X\psi\phi}(Q^2)=g^{\prime}_1\,e^{-g^{\prime}_2\,Q^2}\, .
\end{align}
In this way, we can estimate the uncertainties related to the extrapolation procedure as well.
The results of the fits are shown as solid and dashed lines in  Fig.~\ref{gS0}, 
 which corresponds to the values: $g_1=1.1$ GeV$^{-1}$, $g_2=1.1$ GeV$^2$, $g_3=1.7$ GeV$^2$ and $g^{\prime}_1=0.75$ GeV$^{-1}$ and $g^{\prime}_2=0.5$ GeV$^{-2}$. 
 The resulting coupling constant is
\begin{align}
\mathcal{G}_{X\psi\phi}\equiv g_{X\psi\phi}(-m_{\phi}^2)\simeq 1.115 \pm 0.085 \,\,\mbox{GeV}^{-1}\, .\label{gvalue}
\end{align}

Using this value, we can now evaluate the decay width for the spin $0$ state to $J/\psi\phi$. This decay width is given by,
\begin{align}
\Gamma=\frac{1}{8\pi}\frac{p(m^2_X,m^2_{\psi},m^2_{\phi})}{m^2_X}\frac{1}{2S_X+1}\sum_\text{pol}|\mathcal{M}|^2\, ,\label{dw}
\end{align}
where $p(m^2_X,m^2_{\psi},m^2_{\phi})$ is the center of mass momentum, $S_X$ is the spin of $X$ and $\mathcal{M}$ the reduced matrix element.
Using the Lagrangian of Eq.~(\ref{x}) to calculate the reduced matrix element, we get
\begin{align}\label{gamma}
\Gamma&=\frac{\mathcal{G}_{X\psi\phi}^2}{8\pi m^2_X} p(m^2_X,m^2_{\psi},m^2_{\phi})
\Big[ m^2_{\psi}m^2_{\phi}+\frac{1}{2}(m^2_X-m^2_{\psi}-m^2_{\phi})^2\Big].
\end{align}
Considering the mass of the scalar $X$ as $\sim 4150$ MeV [which is near the central value given in Eq.~(\ref{M0final})], we get the following value for the decay width to $J/\psi\phi$
\begin{align}\label{gammaS0}
\Gamma = (21 \pm 3)\, \mbox{MeV}.
\end{align}
This result, together with the one for the mass sum rule obtained in the previous section [Eq.~(\ref{M0final})], suggests that the interaction between a $D^*_s$ and a $\bar D^*_s$ produces a $I^G(J^{PC})=0^+(0^{++})$ state  with mass around $4.15$ GeV and a decay width into $J/\psi\phi$ of $(21 \pm 3)$ MeV. If instead of using $M_X\sim 4150$ MeV, we take into account the error associated with the mass found within QCD sum rules for the spin 0 state [Eq.~(\ref{M0final})], we find  $\Gamma\sim (34 \pm 14)$ MeV.

Next, we consider the state with spin 2 and determine its decay width to $J/\psi \phi$. In this case, the phenomenological side of the correlation function, after projecting on spin 2, is given by
\begin{align}\label{phenS2}
\Pi_\text{phenom}^{2}&=-\frac{f_{\psi}f_{\phi}\lambda_2 F_{X\psi\phi}(q^2)}{6 m_{\psi} m_{\phi} p^4 ({p^{\prime}}^2-m^2_{\psi})(p^2-m^2_X)(q^2-m^2_{\phi})}\Big\{ p^4 [10 m^2_{\psi}(3m^2_{\phi}-q^2)+p^{\prime 2}(3q^2-10m^2_{\phi})\nonumber\\
&+(p^{\prime}\cdot q)^2]+p^2[(p\cdot q)^2(10m^2_{\psi}-3p^{\prime 2})+(p\cdot p^{\prime})^2(10m^2_{\phi}-3q^2)
-2(p\cdot p^{\prime})(p\cdot q)(p^{\prime}\cdot q)]\nonumber\\
&+4(p\cdot p^{\prime})^2(p\cdot q)^2\Big\}\, .
\end{align}
The procedure to arrive to this expression is analogous to the one followed for the spin 0 case, with the difference that now, when inserting complete sets in Eq.~(\ref{3point}), the current associated with a $X$ state of spin 2, i.e., a tensor, satisfies~\cite{mknno2, Aliev2010,Alhendi2015}
\begin{align}
\langle 0| j_{\alpha\beta} |X(p) \rangle= \lambda_2\epsilon_{\alpha\beta}(p).\label{jj}
\end{align}
In Eq.~(\ref{jj}) $\epsilon_{\alpha\beta} (p)$ is the polarization vector of the tensor field~\cite{Aliev2010,Alhendi2015,Li2015}.
The value for the coupling $\lambda_2$ is given in Eq.~\eqref{M2final}. The form factor present in Eq.~(\ref{phenS2}), $F_{X\psi\phi}(q^2)$, is a consequence of the fact that now the matrix element $X \to  J/\psi $ is given by~\cite{Aliev2010,Alhendi2015}
\begin{align}
\mathcal{M}_{XJ\psi\phi}= F_{X\psi\phi}(q^2) \epsilon_{\rho\sigma}(p)\epsilon^{\rho}(p^{\prime})\epsilon^{\sigma}(q).\label{onemore}
\end{align}
This form factor is related to the following Lagrangian for the $XJ/\psi\phi$ vertex \cite{Giacosa:2005bw}
\begin{align}
\mathcal{L}=i\mathcal{F}_{X\psi\phi}X_{\mu\nu}\psi^\mu\phi^\nu,\label{oL}
\end{align}
which describes the coupling between a tensor state ($X$) and two vector mesons ($J/\psi$ and $\phi$) with a coupling constant $\mathcal{F}_{X\psi\phi}$, which is given in terms of $F_{X\psi\phi}$.  Notice that in Eq.~(\ref{oL}) we describe the vector mesons by vector fields while in Eq.~(\ref{x})  we consider tensor fields. Certainly, there can be more than one way to describe a scalar/tensor meson decay.  Our choice is motivated by the current used in the QCD side and by the method of projecting it to different spins. We must treat both the QCD and the phenomenological side equally, in other words, use the same spin projectors on both sides. Equation~(\ref{x}) corresponds to the decay of a scalar to two vectors and Eq.~(\ref{oL}) describes the decay of a tensor state to two vectors. The tensor field description for $J/\psi$ and $\phi$ in Eq.~(\ref{x}) and vector field description for $J/\psi$ and $\phi$  in Eq.~(\ref{oL})  are suitable for the application  of the projectors of Eq.~(\ref{proj}) to select the desired spin (0 or  2) configuration for the final state.

Analogously to the spin 0 case, the OPE side of Eq.~\eqref{3point} for spin $2$ is determined at leading order in $\alpha_s$ and considering contributions from condensates up to dimension $5$ in the expansion.~We also use
the approximation $p^2\simeq {p^\prime}^2=-P^2$ and perform a Borel transformation on $P^2\to M^2_B$. The result found is given by 
\begin{align}\label{sumrule_s2}
F_{X\psi\phi}(Q^2)C(M^2_B,Q^2)+D\, e^{-u_0/M^2_B}=\int\limits_{4m^2_c}^{s_0}\, ds\, e^{-s/M^2_B}\rho_\text{OPE}^{2}(s,Q^2).
\end{align}
In Eq.~(\ref{sumrule_s2}), $u_0$ and $s_0$ have the same meaning as in the spin 0 case, $D$ represents pole-continuum and continuum-continuum contributions and 
\begin{align}
C(M^2_B,Q^2)&=\frac{f_{\psi}f_{\phi}\lambda^X_2}{24m^5_{\psi}m_{\phi}m^4_{X}(m^2_X-m^2_{\psi})M^2_B(Q^2+m^2_{\phi})}\Big\{ 
M^2_B m^4_X e^{-m^2_{\psi}/M^2_B}[(4m^2_{\psi}Q^2+Q^4)(10m^4_{\psi}\nonumber\\
&+4m^2_{\psi}Q^2+Q^4)+10m^2_{\psi}m^2_{\phi}(12m^4_{\psi}+4m^2_{\psi}Q^2+Q^4)] - Q^4 (m^2_X-m^2_{\psi})[M^2_B(8m^2_Xm^2_{\psi}\nonumber\\
&\times Q^2+10m^2_{\psi}(m^2_{\phi}+m^2_{\psi})m^2_X+Q^4m^2_X+m^2_{\psi}Q^4)-m^2_{\psi}m^2_XQ^4]\nonumber\\
&-M^2_Bm^4_{\psi}e^{-m^2_X/M^2_B}[8m^4_X(5m^2_{\psi}Q^2+2Q^4+15m^2_{\psi}m^2_{\phi}\nonumber\\
&+5m^2_{\phi}Q^2)+2m^2_XQ^4(5m^2_{\psi}+5m^2_{\phi}+4Q^2)+Q^8] \Big\}\, .
\end{align}
The result for each of the contributions to $\rho_\text{OPE}^{2}(s,Q^2)$ can be found in the Appendix B.

\begin{figure}[h!]
\includegraphics[width=0.65\textwidth]{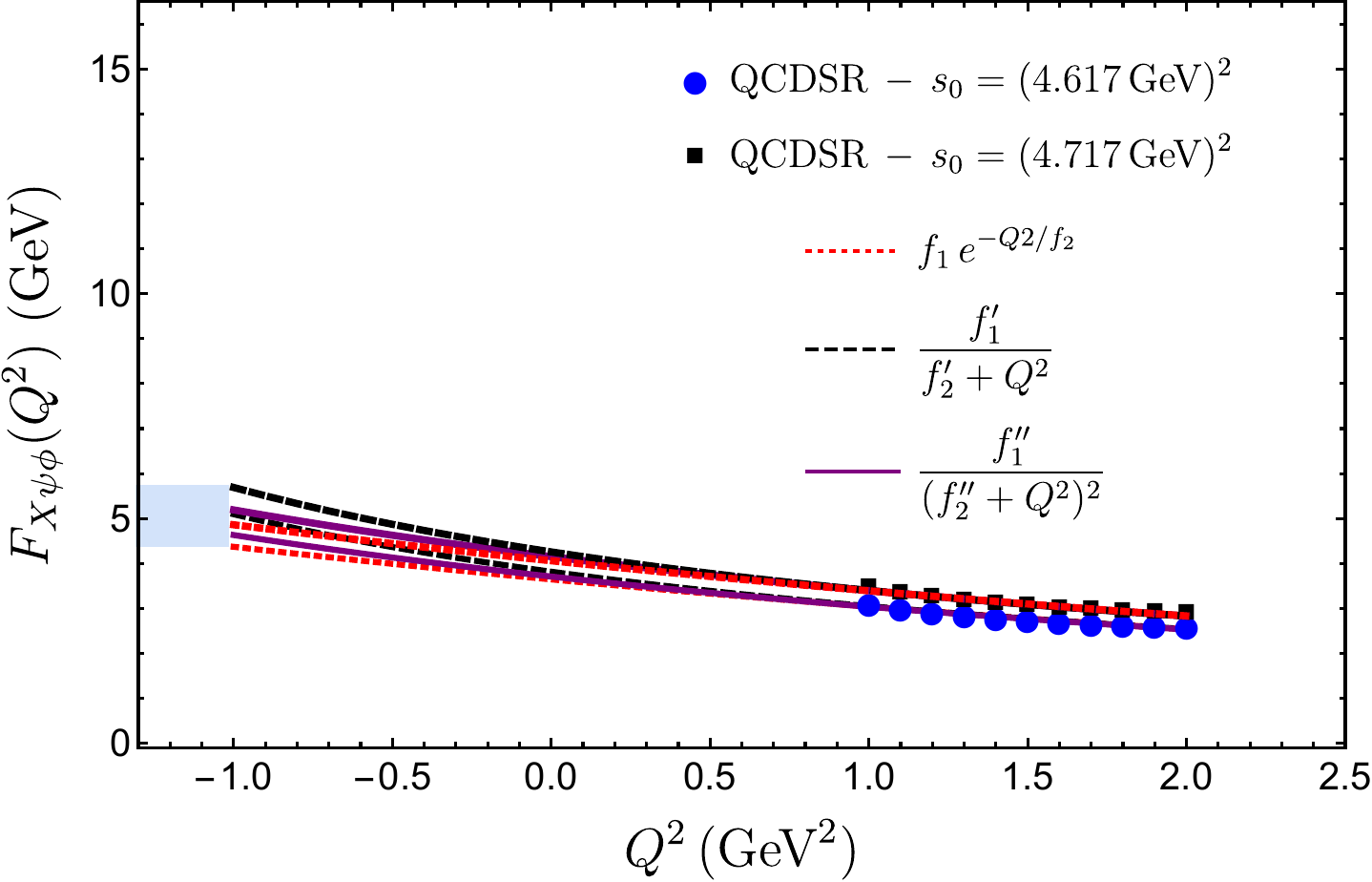}
\caption{The filled circles and boxes in the $Q^2 >0$ region represent the QCD sum rule calculation for $F_{X\psi \phi}$ as a function of  $Q^2$ for two different values of the continuum threshold $s_0$. The value of the Borel mass  $M^2_B$ is fixed and  lies in the region of stability of the sum rules (as shown in Fig.~\ref{g3DS2}). The dotted, dashed and solid lines, for each value of $s_0$, correspond to the results obtained using the form factor given in Eq.~\eqref{ffs2a}-\eqref{ffs2c}. The shaded region near the vertical axis indicates the range of the values for the coupling constant $\mathcal{F}_{X\psi\phi}$.}\label{gS2}
\end{figure}

Using Eq.~(\ref{sumrule_s2}) and its derivative with respect to $1/M^2_B$, we can eliminate $D$ and find an expression for $F_{X\psi\phi}(Q^2)$ which can be calculated numerically.  We show the result found  (as filled circles and boxes)  in Fig.~\ref{gS2} as a function of $Q^2$ for a Borel mass value belonging to the stable sum rule region (shown in Fig~\ref{g3DS2}).
As in the spin 0 case, to extract the coupling constant for the vertex $XJ/\psi \phi$, we need to extrapolate the results obtained for the form factor within the QCD sum rule approach  to a value of $Q^2=-m^2_\phi$. In this case, too, we consider different parameterizations,
\begin{align}\label{ffs2a}
F_{X\psi \phi}(Q^2)=f_1\,e^{-Q^2/f_2}\, ,\\\label{ffs2b}
F_{X\psi \phi}(Q^2)=\frac{f^{\prime}_1}{(f^{\prime}_2+Q^2)}\, ,\\\label{ffs2c}
F_{X\psi \phi}(Q^2)=\frac{f^{\prime\prime}_1}{(f^{\prime\prime}_2+Q^2)^2}\, ,
\end{align}
to fit the results found within the sum rule. The values of the parameters in Eqs.~(\ref{ffs2a})-(\ref{ffs2c}) which fit well the sum rule results are given in Table~\ref{paravalues}. As can be seen in Fig.~\ref{gS2}, all the different expressions  of Eqs.~(\ref{ffs2a})-(\ref{ffs2c}) well represent the QCD sum rule results. 
\begin{table}
\caption{Values of the parameters appearing in Eqs.~(\ref{ffs2a})-(\ref{ffs2c}) for the different values of the continuum threshold $s_0$.}\label{paravalues}
\begin{tabular}{c|c|c}
\hline
Parameter &  $s_0 = (4.617 {\rm GeV})^2$ &$s_0 = (4.717 {\rm GeV})^2$ \\
\hline\hline
$f_1$ & 3.64 GeV & 4.06 GeV\\
$f_2$& 0.18 GeV$^{2}$&0.18 GeV$^{2}$\\
$f^{\prime}_1$ & 15.5 GeV$^3$ & 16.9 GeV$^3$\\
$f^{\prime}_2$ & 3.98 GeV$^2$ &3.97 GeV$^2$\\
$f^{\prime\prime}_1$ & 334.7 GeV$^5$ &371.5 GeV$^5$\\
$f^{\prime\prime}_2$ & 9.5 GeV$^2$ & 9.5 GeV$^2$\\
\hline
\end{tabular}
\end{table}

Taking the result obtained from all three extrapolations, for both values of $s_0$, we find that the coupling constant, $\mathcal{F}_{X\psi\phi}$, for the $XJ/\psi\phi$ vertex is
\begin{align}
\mathcal{F}_{X\psi\phi}\equiv F_{X\psi \phi}(-m^2_{\phi})\simeq 5.0 \pm 0.6 ~\text{GeV} \, ,\label{otr}
\end{align}
which corresponds to the projection on the vertical axis, indicated by the shaded region in Fig.~\ref{gS2}.

From the Lagrangian of Eq.~(\ref{oL}), we can determine the reduced matrix element present in Eq.~(\ref{dw}), and calculate in this way the decay width of the state with spin 2 to $J/\psi\phi$. 
We get
\begin{align}
\sum_\text{pol}|\mathcal{M}|^2&=\frac{(\mathcal{F}_{X\psi\phi})^2}{24m^4_Xm^2_{\psi}m^2_{\phi}}\Big\{ m^8_X+6m^6_X(m^2_{\psi}+m^2_{\phi})-14m^4_X(m^4_{\psi}+m^4_{\phi}-6m^2_{\psi}m^2_{\phi})\nonumber\\
&+6m^2_X(m^2_{\phi}-m^2_{\psi})^2(m^2_{\psi}+m^2_{\phi})+(m^2_{\phi}-m^2_{\psi})^4\Big\}\, .
\end{align}
and considering $m_X\simeq 4150$ MeV, which is a value close to the central one given in Eq.~(\ref{M2final}), we obtain
\begin{align}\label{gamma_S2}
 \Gamma = (13 \pm 2 )\,\,\mbox{MeV}\, .
\end{align}
Thus, we find that the interaction between a $D^*_s$ and a $\bar{D}^*_s$ in spin 2 produces a $I^G(J^{PC})=0^+(2^{++})$ state with mass around 4.15 GeV and a decay width to $J/\psi \phi$ of ($13 \pm 2$) MeV. We can also estimate the change in the width due to the error associated with the mass of the spin 2 state found within QCD sum rules [Eq.~(\ref{M2final})], finding that  $\Gamma = ( 20 \pm 7)$ MeV.

From our study, both $J^{PC} =$  0$^{++}$ and 2$^{++}$ are attributable to $X(4160)$. At present, there is no experimental data on the decay width of $X(4160)$
to the channel $J/\psi\phi$, but we can compare our results with those evaluated using the findings of Ref.~\cite{raquel}. In Ref.~\cite{raquel}, the authors study the $D^*_s\bar D^*_s$ system using effective field theories and they arrive to the conclusion that the interaction between these two particles in spin 2 generates a state with a mass 4170 MeV and full width of 130 MeV. Since the full width found is compatible with the one found for $X(4160)$, the authors of Ref. ~\cite{raquel} associate their $I^G(J^{PC})=0^+(2^{++})$ state with $X(4160)$. Although the partial decay widths to the different channels are not given in Ref.~\cite{raquel}, the coupling of the state to different channels is provided (which corresponds to the residues of the scattering matrix at the pole position). Since
$J/\psi\phi$ is one of these channels, it is possible to calculate the partial decay width to $J/\psi\phi$ considering the coupling given by the authors and assuming a Breit-Wigner form for the scattering matrix near the pole position as done in Ref.~\cite{oller}.
Following Ref.~\cite{oller} and using the coupling of the state to $J/\psi\phi$ of Ref.~\cite{raquel}, we have calculated the partial decay width of the $0^+(2^{++})$ state of Ref.~\cite{raquel} to $J/\psi \phi$, obtaining the result $\sim 20-30$ MeV 
Thus, our result
for the decay width of the  $0^+(2^{++})$ found using QCD sum rules is compatible with the one obtained from the model of Ref.~\cite{raquel}.

The comparison made above hints a possible $D^*_s\bar D^*_s$ molecule-like nature with quantum numbers $J^{PC} = 2^{++}$ for $X(4160)$. However, our work also implies the existence of a $J^{PC} = 0^{++}$ state with a similar mass and partial decay width to $J/\psi \phi$. Interestingly, although the analysis of the recent $J/\psi \phi$ mass distribution made in Ref.~\cite{lhcb}  concludes the presence of one resonance around 4140 MeV with preferred $J^{PC} = 1^{++}$, fits made with quantum numbers $0^{++}, 2^{++}$ also have good statistical significance. A fit to the data of Ref.~\cite{lhcb} with more than one resonance present around 4100 MeV has not been studied. Our findings hint that more than one resonance may contribute  around 4100 MeV in the data of Ref.~\cite{lhcb}. It would be interesting to test this information in future investigations.

\section{Summary and Conclusions}

We can conclude the present work by stating that a study of  the $\bar D^*_s D^*_s$ system has been made using QCD sum rules and three states with $I^G=0^+$, $J^{PC} = 0^{++}$, $1^{+-}$, $2^{++}$ and mass around 4.12 GeV have been found, with an error bar for the mass of $\simeq$ $\pm 0.1$ GeV. In spite of having now currents with strange quarks, these results, considering the error bars, are compatible with those found in our previous study of the $\bar D^* D^*$ system, where states with mass around 3.9 $\pm$ 0.1 GeV were found  with $J^{PC} = 0^{++}$, $1^{+-}$, $2^{++}$. This raises the question if the states found with $\bar D^* D^*$ and $\bar D^*_s D^*_s$ current are same or not. A way to disentangle these states is the calculation of their decay width into the $J/\psi\phi$ channel, since decay of the $D^* \bar D^*$ states to such a channel is OZI suppressed and expected to be small. As a result we have found that the $0^{++}$ and $2^{++}$ states obtained using the $D^*_s\bar D^*_s$ current have  similar partial decay width to $J/\psi\phi$. We find that our results are compatible with the findings of other theoretical works. For example, in Ref.~\cite{raquel} the $\bar D^* D^*$, $\bar D^*_s D^*_s$ systems were studied using a coupled channel formalism within effective field theories and the authors found three states with a mass around 3900 MeV with quantum numbers $I^G (J^{PC})=0^+ (0^{++})$, $0^- (1^{+-})$, $0^+ (2^{++})$ and one state with mass  around 4.1 GeV with $J^{PC} = 2^{++}$. Yet another study within an effective field theory approach based on heavy quark symmetry \cite{HidalgoDuque:2012pq} finds three states with a mass around 3900 MeV with quantum numbers $I^G (J^{PC})=0^+ (0^{++})$, $0^- (1^{+-})$, $0^+ (2^{++})$ and two states with mass around 4.1 GeV and $J^{PC} = 0^{++}$ and $1^{+-}$.  
\\
\\
\noindent
{\bf Acknowledgement:} The authors thank the financial support of FAPESP and CNPq.
\\
\\
\noindent
\appendix{\bf{Appendix A}}

We present here the results for the spectral density in the OPE series corresponding to the two point correlation function.  We define the following functions
\begin{align}
F &\equiv m_c^2 \left( \alpha +\beta \right) - q^2 \alpha \beta,\quad \tilde{g} \equiv  1- \alpha -\beta, \nonumber\\
h &\equiv  q^2  \alpha \beta, \quad \mathcal{F}\equiv \frac{m^2_c \left(\eta + \xi \right)}{\eta \xi},\nonumber\\
\mathcal{G} &\equiv 1- \eta - \xi,
\end{align}
where $\alpha,\beta, \eta, \xi$ are variables of integration, $m_c$ is the constituent mass of the charm quark, and $q$ is the running momentum in the 
correlation function.

\begin{align}
\rho^{0}_{\textrm{pert}}&=\frac{1}{2^{12}\pi^6}\mathlarger{\int}\limits_{\alpha_\text{min}}^{\alpha_\text{max}} \frac{d\alpha}{\alpha^3} \mathlarger{\int}\limits_{\beta_\text{min}}^{\beta_\text{max}}\frac{d\beta}{\beta^3} F^2\tilde{g} \left[F^2(\tilde{g}^2-2\tilde{g}+6)-16F(\tilde{g}-1)h+16\tilde{g}^2h^2\right]\nonumber\\
\rho^{1}_{\textrm{pert}}&=\frac{3}{2^{11}\pi^6} \mathlarger{\int}\limits_{\alpha_\text{min}}^{\alpha_\text{max}} \frac{d\alpha}{\alpha^3} \mathlarger{\int}\limits_{\beta_\text{min}}^{\beta_\text{max}}\frac{d\beta}{\beta^3}
 F^3 \tilde{g} \left[F(1-\tilde{g})+8 \tilde{g} h\right]\nonumber\\
 \rho^{2}_{\textrm{pert}}&=\frac{1}{2^{11}\pi^6} \mathlarger{\int}\limits_{\alpha_\text{min}}^{\alpha_\text{max}} \frac{d\alpha}{\alpha^3} \mathlarger{\int}\limits_{\beta_\text{min}}^{\beta_\text{max}}\frac{d\beta}{\beta^3}F^2 \tilde{g} \left[F^2 \left(\tilde{g}^2+\tilde{g}+3\right)-8 F \tilde{g} (2 \tilde{g}+1) h+16 \tilde{g}^2 h^2\right]\nonumber\\ 
\rho^{0}_{m_s}&=-\frac{m_c m_s}{2^{10}\pi^6}\mathlarger{\int}\limits_{\alpha_\text{min}}^{\alpha_\text{max}} \frac{d\alpha}{\alpha^3} \mathlarger{\int}\limits_{\beta_\text{min}}^{\beta_\text{max}}\frac{d\beta}{\beta^3} F^2 \tilde{g}  (\alpha +\beta ) \left[F (\tilde{g}+2)-6 \tilde{g} h\right] \nonumber\\
\rho^{1}_{m_s}&=-3\rho^{0}_{m_s},\quad\rho^{2}_{m_s}=5\rho^{0}_{m_s}\nonumber
  \end{align}
 \begin{align}
\rho^{0}_{\langle\bar s s\rangle}&= -\frac{ m_c \langle\bar s s\rangle}{2^7 \pi^4}\mathlarger{\int}\limits_{\alpha_\text{min}}^{\alpha_\text{max}} \frac{d\alpha}{\alpha^2} \mathlarger{\int}\limits_{\beta_\text{min}}^{\beta_\text{max}}\frac{d\beta}{\beta^2} F\left[\alpha +\beta\right]\left[\tilde{g}\left(F - 4 h\right)  + F\right]\nonumber\\
\rho^{1}_{\langle\bar s s\rangle}&= -3\rho^{0}_{\langle\bar s s\rangle},\quad \rho^{2}_{\langle\bar s s\rangle}= 5\rho^{0}_{\langle\bar s s\rangle}\nonumber\\
\rho^{0}_{m_s\langle\bar s s\rangle}&=\frac{m_s\left\langle s \bar{s}\right\rangle}{2^{7}\pi^4}\left\{\mathlarger{\int}\limits_{\alpha_\text{min}}^{\alpha_\text{max}} \frac{d\alpha}{\alpha} \mathlarger{\int}\limits_{\beta_\text{min}}^{\beta_\text{max}}\frac{d\beta}{\beta}   \left[F^2 (3 \tilde{g}-2)+F \{(8-24 \tilde{g}) h+12m^2_c\}+8 \tilde{g} h^2\right]\right.\nonumber\\
&\left.\quad-3\mathlarger{\int}\limits_{\alpha_\text{min}}^{\alpha_\text{max}} d\alpha\frac{\left[m_c^2+\alpha(\alpha-1)q^2\right]^2}{\alpha(\alpha-1)}\right\}\nonumber\\
\rho^{1}_{m_s\langle\bar s s\rangle}&=-\frac{3m_s\left\langle s \bar{s}\right\rangle}{2^{7}\pi^4}\left\{2\mathlarger{\int}\limits_{\alpha_\text{min}}^{\alpha_\text{max}} \frac{d\alpha}{\alpha} \mathlarger{\int}\limits_{\beta_\text{min}}^{\beta_\text{max}}\frac{d\beta}{\beta}F(F-4 h+6m^2_c)+\mathlarger{\int}\limits_{\alpha_\text{min}}^{\alpha_\text{min}} d\alpha\frac{\left[m_c^2+\alpha(\alpha-1)q^2\right]^2}{\alpha(\alpha-1)}\right\}\nonumber\\
\rho^{2}_{m_s\langle\bar s s\rangle}&=\frac{m_s\left\langle s \bar{s}\right\rangle}{2^{7}\pi^4}\left\{2\mathlarger{\int}\limits_{\alpha_\text{min}}^{\alpha_\text{max}} \frac{d\alpha}{\alpha} \mathlarger{\int}\limits_{\beta_\text{min}}^{\beta_\text{max}}\frac{d\beta}{\beta}\left[F^2 (3 \tilde{g}+1)+F \{30m^2_c-4 (6 \tilde{g}+1) h\}+8 \tilde{g} h^2\right]\right.\nonumber\\
&\quad\left.-3\mathlarger{\int}\limits_{\alpha_\text{min}}^{\alpha_\text{max}} d\alpha\frac{\left[m_c^2+\alpha(\alpha-1)q^2\right]^2}{\alpha(\alpha-1)}\right\}\nonumber\\
\rho^{0}_{\langle g^2 G^2\rangle}&= \frac{\langle g^2 G^2\rangle}{3\cdot2^{13}\pi^6}\left\{\mathlarger{\int}\limits_{\alpha_\text{min}}^{\alpha_\text{max}} \frac{d\alpha}{\alpha^3} \mathlarger{\int}\limits_{\beta_\text{min}}^{\beta_\text{max}}\frac{d\beta}{\beta^3}(\alpha +\beta )\left[2 m_c^2 \tilde{g} \left(\alpha ^2-\alpha  \beta +\beta ^2\right)  \left(F \left\{\tilde{g}^2+6\right\}+2 \tilde{g}h \{3-2 \tilde{g}\}\right)\right.\right.\nonumber\\
&\quad\left.\left.-4 \tilde{g}^2 \left(\alpha ^3+\beta ^3\right) m_c^4+\alpha  \beta  \left(F^2 \left\{3 \tilde{g}^2-8 \tilde{g}+2\right\}-8 F \tilde{g} \{3 \tilde{g}-4\} h+8 \tilde{g}^2 h^2\right)\right]\right.\nonumber\\
&\left.\quad+\frac{8m_c^6}{3}\mathlarger{\int}\limits_{0}^{1} \frac{d\eta}{\eta^4} \mathlarger{\int}\limits_{0}^{1-\eta} \frac{d\xi}{\xi^4} \mathcal{G}^3 (\mathcal{G} -1)^3 \left(\eta ^2-\eta  \xi +\xi ^2\right) \delta(s-\mathcal{F})\right\}\nonumber\\
\rho^{1}_{\langle g^2 G^2\rangle}&= \frac{\langle g^2G^2 \rangle }{2^{12}\pi^6}  \mathlarger{\int}\limits_{\alpha_\text{min}}^{\alpha_\text{max}} \frac{d\alpha}{\alpha^3} \mathlarger{\int}\limits_{\beta_\text{min}}^{\beta_\text{max}}\frac{d\beta}{\beta^3} (1-\tilde{g})\left[2 \tilde{g} m_c^2(\alpha ^2-\alpha  \beta +\beta ^2) (F+3 \tilde{g} h)-2 \tilde{g}^2 m_c^4(\alpha ^3+\beta ^3) +3 \alpha  \beta  F^2\right]\nonumber\\
\rho^{2}_{\langle g^2 G^2\rangle}&= \frac{\langle g^2G^2 \rangle }{3\cdot 2^{12}\pi^6}\left\{\mathlarger{\int}\limits_{\alpha_\text{min}}^{\alpha_\text{max}} \frac{d\alpha}{\alpha^3} 
\mathlarger{\int}\limits_{\beta_\text{min}}^{\beta_\text{max}}\frac{d\beta}{\beta^3}(\alpha +\beta) \left[2 \tilde{g}m_c^2(\alpha ^2-\alpha  \beta +\beta ^2)  \left\{F (\tilde{g}^2+3)-\tilde{g} (4 \tilde{g}+3) h\right\}\right.\right.\nonumber\\
&\left.\left.\quad+2 \tilde{g}^2 (\alpha ^3+\beta ^3) m_c^4+\alpha  \beta \left\{F^2 (3 \tilde{g}^2-8 \tilde{g}-7)-8 F \tilde{g} (3 \tilde{g}-4) h+8 \tilde{g}^2 h^2\right\}\right]\right\}\nonumber
  \end{align}
 \begin{align}
\rho^{0}_{\langle\bar s g\sigma G s\rangle}&= -\frac{m_c \langle\bar s g\sigma \cdot G s\rangle}{2^8 \pi^4} \mathlarger{\int}\limits_{\alpha_\text{min}}^{\alpha_\text{max}} \frac{d\alpha}{\alpha}
\left\{\frac{ m_c^2 - q^2 \alpha \left(1- \alpha\right)}{1- \alpha} +\mathlarger{\int}\limits_{\beta_\text{min}}^{\beta_\text{max}}\frac{d\beta}{\beta} \left[m_c^2\left(1-\tilde{g} \right) - 3 h\right] \left( 1-\tilde{g}\right) \right\}\nonumber\\
\rho^{1}_{\langle\bar s g\sigma G s\rangle}&=-3 \rho^{0}_{\langle\bar s g\sigma  G s\rangle},\quad
\rho^{2}_{\langle\bar s g\sigma G s\rangle}=5 \rho^{0}_{\langle\bar s g\sigma  G s\rangle},\nonumber\\
\rho^{0}_{m_s\langle\bar s g\sigma G s\rangle}&= \frac{m_s\langle\bar s g\sigma \cdot G s\rangle}{3\cdot 2^8  \pi^4} \left\{\mathlarger{\int}\limits_{\alpha_\text{min}}^{\alpha_\text{max}} \frac{d\alpha}{\alpha}
\mathlarger{\int}\limits_{\beta_\text{min}}^{\beta_\text{max}} \frac{d\beta}{\beta}\left[4 m_c^2 (\alpha +\beta )^2 -3 (F \{\tilde{g} (\alpha +\beta )-2 \alpha  \beta \}\right.\right.\nonumber\\
&\quad\left.\left.+4 h \{2 \alpha  \beta +\alpha +\beta -\alpha  \tilde{g}-\beta  \tilde{g}\})\right]\right.\nonumber\\
&\quad\left.+\mathlarger{\int}\limits_{\alpha_\text{min}}^{\alpha_\text{max}} d\alpha\frac{m_c^2\left(26 \alpha ^2-26 \alpha +1\right) +\left(6 \alpha ^3-12 \alpha ^2+7 \alpha -1\right) \alpha  q^2}{(\alpha-1)\alpha}\right.\nonumber\\
&\quad\left.+4m^4_c\mathlarger{\int}\limits_{0}^{1} \frac{d\eta}{\eta^2}\mathlarger{\int}\limits_{0}^{1-\eta} \frac{d\xi}{\xi^2} (\eta +\xi )^2 [2 \eta  \xi -\mathcal{G} (\eta +\xi )] \delta(s-\mathcal{F})\right\},\nonumber\\
\rho^{1}_{m_s\langle\bar s g\sigma G s\rangle}&=-\frac{3m_s\langle\bar s g\sigma \cdot G s\rangle}{2^8 \pi^4} \mathlarger{\int}\limits_{\alpha_\text{min}}^{\alpha_\text{max}} d\alpha\frac{m^2_c(6 \alpha ^2-6 \alpha -1)+
(2 \alpha ^3-4 \alpha ^2+\alpha +1) \alpha  q^2}{(\alpha -1) \alpha}\nonumber\\
\rho^{2}_{m_s\langle\bar s g\sigma G s\rangle}&=\frac{m_s\langle\bar s g\sigma \cdot G s\rangle}{3\cdot 2^8 \pi^4}\left\{ 2\mathlarger{\int}\limits_{\alpha_\text{min}}^{\alpha_\text{max}} \frac{d\alpha}{\alpha}
\mathlarger{\int}\limits_{\beta_\text{min}}^{\beta_\text{max}}\frac{d\beta}{\beta}\left[4m_c^2 (\alpha +\beta )^2\right.\right.\nonumber\\
&\quad\left.\left.-3 (F \{\tilde{g} (\alpha +\beta )-2 \alpha  \beta \}+4 h \{2 \alpha  \beta +\alpha +\beta -\alpha  \tilde{g}-\beta  \tilde{g}\})\right]\right.\nonumber\\
&\left.+\mathlarger{\int}\limits_{\alpha_\text{min}}^{\alpha_\text{max}} d\alpha\frac{m_c^2(106 \alpha ^2-106 \alpha -7) +(30 \alpha ^3-60 \alpha ^2+23 \alpha +7) \alpha  q^2 }{(\alpha-1)\alpha}\right.\nonumber\\
&\left.-8m_c^4\mathlarger{\int}\limits_{0}^{1} d\eta\mathlarger{\int}\limits_{0}^{1-\eta} d\xi  (\eta +\xi )^2 [2 \eta  \xi -\mathcal{G} (\eta +\xi )] \delta(s-\mathcal{F})\right\}\nonumber
  \end{align}
 \begin{align}
\rho^{0}_{{\langle\bar s s\rangle}^2}&=\mathlarger{\int}\limits_{\alpha_\text{min}}^{\alpha_\text{max}} d\alpha \frac{m_c^2 \langle\bar s s\rangle^2}{2^4 \pi^2},\quad
\rho^{1}_{{\langle\bar s s\rangle}^2}=-3 \rho^{0}_{{\langle\bar s s\rangle}^2},\quad\rho^{2}_{{\langle\bar s s\rangle}^2}=5\rho^{0}_{{\langle\bar s s\rangle}^2}\nonumber\\
\rho^{0}_{m_s\langle\bar s s\rangle^2}&=-\frac{m_c m_s\langle\bar s s\rangle^2}{2^5 \pi^2}\left\{\mathlarger{\int}\limits_{\alpha_\text{min}}^{\alpha_\text{max}} d\alpha -m^2_c\mathlarger{\int}\limits_{0}^{1} d\eta\frac{1}{\eta(\eta-1)}\delta\left[s-\frac{m^2_c}{\eta(1-\eta)}\right]\right\},\nonumber\\
\rho^{1}_{m_s\langle\bar s s\rangle^2}&=-3\rho^{0}_{m_s\langle\bar s s\rangle^2},\quad\rho^{2}_{m_s\langle\bar s s\rangle^2}=5\rho^{0}_{m_s\langle\bar s s\rangle^2}\nonumber\\
\rho^{0}_{\langle g^3 G^3\rangle}&= \frac{\langle g^3G^3 \rangle }{3\cdot 2^{14}\pi^6} \left\{\mathlarger{\int}\limits_{\alpha_\text{min}}^{\alpha_\text{max}} \frac{d\alpha}{\alpha^3}
\mathlarger{\int}\limits_{\beta_\text{min}}^{\beta_\text{max}}\frac{d\beta}{\beta^3}\tilde{g}\left[2 m_c^2 \{\alpha^4 (\tilde{g}^2-3 \tilde{g}+6)+\beta ^4 (\tilde{g}^2-3 \tilde{g}+6)-\alpha ^3 \beta  \tilde{g}-\alpha  \beta ^3 \tilde{g}\}\right.\right.\nonumber\\
&\left.\left.\quad+(\alpha ^3+\beta ^3)\{F (\tilde{g}^2+6)+2 \tilde{g}h (3-2 \tilde{g}) \}\right]\right.\nonumber\\
&\quad\left.+\frac{4 m^4_c}{3M^2_B}\mathlarger{\int}\limits_{0}^{1} \frac{d\eta}{\eta^5}\mathlarger{\int}\limits_{0}^{1-\eta}\frac{d\xi}{\xi^5} \mathcal{G}^2  (\eta +\xi )\left[\mathcal{G} \left\{\eta  M^2_B \xi (\eta ^2+\eta  \xi +\xi ^2) (\eta -\xi )^2\right.\right.\right.\nonumber\\
&\quad\Bigg.\left.\left.+2 m^2_c(\eta ^5+\eta ^4 \xi +\eta  \xi ^4+\xi ^5)\right\}-6 \eta  M^2_B \xi  (\eta ^4+\xi ^4)\delta(s-\mathcal{F})\right]\Bigg\}\nonumber\\
\rho^{1}_{\langle g^3 G^3\rangle}&= \frac{\langle g^3G^3 \rangle }{2^{13}\pi^6}\left\{ \mathlarger{\int}\limits_{\alpha_\text{min}}^{\alpha_\text{max}} \frac{d\alpha}{\alpha^3}
\mathlarger{\int}\limits_{\beta_\text{min}}^{\beta_\text{max}}\frac{d\beta}{\beta^3}g\left[(\alpha^3+\beta^3) (F+3 \tilde{g} h)-m_c^2 \{\alpha ^4 (3 \tilde{g}-2)+\alpha ^3 \beta  \tilde{g}+\alpha  \beta ^3 \tilde{g}\right.\right.\nonumber\\
&\quad\left.\left.+\beta ^4 (3 \tilde{g}-2)\}\right] -4m_c^4\mathlarger{\int}\limits_{0}^{1} \frac{d\eta}{\eta^4}\mathlarger{\int}\limits_{0}^{1-\eta}\frac{d\xi}{\xi^4}\mathcal{G}^2  (\eta +\xi ) (\eta ^4+\xi ^4)\delta(s-\mathcal{F})\right\}\nonumber\\
\rho^{2}_{\langle g^3 G^3\rangle}&=\frac{\langle g^3G^3 \rangle }{9\cdot 2^{13}\pi^6} \left\{3\mathlarger{\int}\limits_{\alpha_\text{min}}^{\alpha_\text{max}} \frac{d\alpha}{\alpha^3}
\mathlarger{\int}\limits_{\beta_\text{min}}^{\beta_\text{max}}\frac{d\beta}{\beta^3}\tilde{g} \left[m_c^2 (\alpha^4 \{2 \tilde{g}^2+3 \tilde{g}+6\}+\beta^4 \{2 \tilde{g}^2+3 \tilde{g}+6\}+\alpha^3 \beta  \tilde{g}+\alpha  \beta^3 \tilde{g})\right.\right.\nonumber\\
&\left.\quad+(\alpha^3+\beta^3) (F \{\tilde{g}^2+3\}-\tilde{g} \{4 \tilde{g}+3\} h)\right]\nonumber\\
&\quad+\frac{4m_c^4}{M^2_B}\mathlarger{\int}\limits_{0}^{1} \frac{d\eta}{\eta^5}\mathlarger{\int}\limits_{0}^{1-\eta}\frac{d\xi}{\xi^5}\mathcal{G}^2  (\eta +\xi )\left(\mathcal{G} \left[\eta  M_B^2 \xi  \left(\eta^2+\eta  \xi +\xi^2\right) \left(\eta -\xi \right)^2\right.\right.\nonumber\\
&\quad\left.\left.\left.+2 m_c^2 \left(\eta ^5+\eta ^4 \xi +\eta  \xi ^4+\xi ^5\right)\right]+3 \eta  M_B^2 \xi  \left[\eta^4+\xi ^4\right]\right)\delta(s-\mathcal{F})\right\}\nonumber
\end{align}
\begin{align}
&\rho^{0}_{\langle \bar s s\rangle \langle \bar sg\sigma G s\rangle}=\frac{m^2_c\langle \bar s s\rangle\langle \bar sg\sigma\cdot G s\rangle}{32\pi^2M^2_B}\int_0^1 d\eta \frac{[m^2_c-\eta(\eta-1)M^2_B]}{\eta(\eta-1)}\delta\left[s-\frac{m^2_c}{\eta(1-\eta)}\right]\nonumber\\
&\rho^{1}_{\langle \bar s s\rangle \langle \bar sg\sigma G s\rangle}=-3\rho^{0}_{\langle \bar s s\rangle \langle \bar sg\sigma G s\rangle},\quad\rho^{2}_{\langle \bar s s\rangle \langle \bar sg\sigma G s\rangle}=5\rho^{0}_{\langle \bar s s\rangle \langle \bar sg\sigma G s\rangle}\nonumber
\end{align}
\begin{align}
\rho^{0}_{m_s\langle\bar s s\rangle\langle \bar s g \sigma G s\rangle}&=\frac{m_c m_s\langle\bar s s\rangle\langle \bar s g \sigma \cdot G s\rangle}{3\cdot2^7\pi^2 M^4_B}\Bigg(\int_0^1d\eta \frac{[5m^4_c-9(\eta-1)\eta m^2_cM^2_B+8(\eta-1)^2\eta^2M^4_B]}{\eta^2(\eta-1)^2}\Bigg.\nonumber\\
&\Bigg.\delta\left[s-\frac{m^2_c}{\eta(1-\eta)}\right]\Bigg)\nonumber\\
\rho^{1}_{m_s\langle\bar s s\rangle\langle \bar s g \sigma G s\rangle}&=-3\rho^{0}_{m_s\langle\bar s s\rangle\langle \bar s g \sigma G s\rangle},\quad \rho^{2}_{m_s\langle\bar s s\rangle\langle \bar s g \sigma G s\rangle}=5 \rho^{0}_{m_s\langle\bar s s\rangle\langle \bar s g \sigma G s\rangle}\nonumber
\end{align}

The limits of integration in the above expressions are
\begin{align}
\alpha_\text{min} =\frac{ 1- \sqrt{1 - \frac{4 m_c^2}{q^2}}}{2}, \hspace{0.2cm} \alpha_\text{max} =\frac{ 1+ \sqrt{1 - \frac{4 m_c^2}{q^2}}}{2}, \hspace{0.2cm}
\beta_\text{min} =\frac{ m_c^2 \alpha}{\left(\alpha q^2- m_c^2\right)}, \hspace{0.2cm} \beta_\text{max} = 1 - \alpha,\nonumber
\end{align}
and $M_B$ represents the Borel mass.
\\
\\
\noindent
\appendix{\bf{Appendix B}}

In what follows we present the spectral densities obtained for the three-point function for spin $0$ and $2$ cases.\\
Spin $0$ case:
\begin{align}
\rho^0_\text{pert}(s,Q^2)&=\frac{1}{36 s^3}\sqrt{1-\frac{4m^2_c}{s}}(2m^2_c+s)\Big\{-\frac{1}{8}Q^4 s\Big(2s+Q^2\Big)-\frac{1}{4}\Big(2s+Q^2\Big)^2\Big(Q^2 s+\frac{Q^4}{2}\Big)\nonumber\\
&+s\Big[\frac{Q^4 s}{4}+s\Big(\frac{Q^4}{4}-Q^2 s\Big)\Big]\Big\}\,\mbox{log}\left(\frac{Q^2}{\Lambda^2_\text{QCD}}\right)\, .\nonumber
\end{align}
\begin{align}
\rho^0_{\ss}(s,Q^2)&=-\frac{\pi^2 m_c \ss}{6 s}\sqrt{1-\frac{4 m^2_c}{s}}(14 s +5Q^2)\, ,\nonumber
\end{align}
\begin{align}
\rho^0_{\mixs}(s,Q^2)&=\frac{\pi^2 m_c \mixs}{18 Q^2 s}\sqrt{1-\frac{4m^2_c}{s}}(14s+5Q^2)\, ,\nonumber
\end{align}

Spin $2$ case:
\begin{align}
\rho^2_\text{pert}(s,Q^2)&=-\frac{1}{36 s^3}\sqrt{1-\frac{4m^2_c}{s}}(2m^2_c+s)\Big\{\frac{Q^4 s}{4}\Big(2 s +Q^2\Big)+\frac{1}{4}\Big(2s+Q^2\Big)^2 \Big(Q^4-7Q^2s\Big)\nonumber\\
&+s\Big[\frac{7Q^4 s}{4}+s\Big(\frac{Q^4}{4}-13Q^2 s\Big)\Big]\Big\}\,\mbox{log}\left(\frac{Q^2}{\Lambda^2_\text{QCD}}\right)\, ,\nonumber
\end{align}
\begin{align}
\rho^2_{\ss}(s,Q^2)&=-\frac{10\pi^2 m_c \ss}{3 s}\sqrt{1-\frac{4 m^2_c}{s}}(3 s +Q^2)\, ,\nonumber
\end{align}
\begin{align}
\rho^2_{\mixs}(s,Q^2)&=-\frac{10\pi^2 m_c \mixs}{9 Q^2 s}\sqrt{1-\frac{4m^2_c}{s}}(s+Q^2)\, ,\nonumber
\end{align}
with $\Lambda_\text{QCD}\sim 0.1-0.25$ GeV.


\newpage
\begin{thebibliography}{99}
\bibitem{godfrey} 
  S.~Godfrey and S.~L.~Olsen,
  Ann.\ Rev.\ Nucl.\ Part.\ Sci.\  {\bf 58}, 51 (2008).
    
 \bibitem{yuan}
 C.~Z.~Yuan [BESIII Collaboration],
  Front.\ Phys.\ China {\bf 10}, 101401 (2015).

\bibitem{liu} 
  X.~Liu,
  Chin.\ Sci.\ Bull.\  {\bf 59}, 3815 (2014).
  

   \bibitem{nielsen} 
  M.~Nielsen, F.~S.~Navarra and S.~H.~Lee,
  Phys.\ Rept.\  {\bf 497}, 41 (2010).
 
 \bibitem{olsen} S. L. Olsen, Front. Phys. 10, 101401 (2015).


 \bibitem{hosaka} 
  A.~Hosaka, T.~Iijima, K.~Miyabayashi, Y.~Sakai and S.~Yasui,
  arXiv:1603.09229 [hep-ph].
  
  \bibitem{Nielsen:2014mva} 
  M.~Nielsen and F.~S.~Navarra,
  Mod.\ Phys.\ Lett.\ A {\bf 29}, 1430005 (2014).
  
  \bibitem{eef1} 
  E.~van Beveren, G.~Rupp and J.~Segovia,
  Phys.\ Rev.\ Lett.\  {\bf 105}, 102001 (2010).
  
  \bibitem{vijande} 
  J.~Vijande and A.~Valcarce,
  Phys.\ Lett.\ B {\bf 736}, 325 (2014).
  
  \bibitem{brodsky} 
  S.~J.~Brodsky, D.~S.~Hwang and R.~F.~Lebed,
  Phys.\ Rev.\ Lett.\  {\bf 113}, 112001 (2014).
  
  \bibitem{braaten} 
  E.~Braaten, C.~Langmack and D.~H.~Smith,
  Phys.\ Rev.\ D {\bf 90}, 014044 (2014)
  
  \bibitem{esposito} 
  A.~Esposito, A.~L.~Guerrieri, F.~Piccinini, A.~Pilloni and A.~D.~Polosa,
  Int.\ J.\ Mod.\ Phys.\ A {\bf 30}, 1530002 (2015).
  
  \bibitem{chen} 
  H.-X.~Chen, L.~Maiani, A.~D.~Polosa and V.~Riquer,
  Eur.\ Phys.\ J.\ C {\bf 75},  550 (2015).
  
  \bibitem{eef2} 
  M.~Cardoso, G.~Rupp and E.~van Beveren,
  Eur.\ Phys.\ J.\ C {\bf 75}, 26 (2015).
  
  \bibitem{ma} 
  L.~Ma, X.~H.~Liu, X.~Liu and S.~L.~Zhu,
  Phys.\ Rev.\ D {\bf 91}, 034032 (2015).
  
  \bibitem{mko1} 
  A.~Martinez Torres, K.~P.~Khemchandani, L.~S.~Geng, M.~Napsuciale and E.~Oset,
  Phys.\ Rev.\ D {\bf 78}, 074031 (2008).
  
  \bibitem{mko2} 
  A.~Martinez Torres, K.~P.~Khemchandani, D.~Gamermann and E.~Oset,
  Phys.\ Rev.\ D {\bf 80}, 094012 (2009).
  
  \bibitem{daniel}
  D.~Gamermann, J.~Nieves, E.~Oset and E.~Ruiz Arriola,
  Phys.\ Rev.\ D {\bf 81}, 014029 (2010).
  
   \bibitem{branz}
   T.~Branz, R.~Molina and E.~Oset,
  Phys.\ Rev.\ D {\bf 83}, 114015 (2011).
  
  
  \bibitem{ortega} 
  P.~G.~Ortega, D.~R.~Entem and F.~Fernandez,
  J.\ Phys.\ G {\bf 40}, 065107 (2013).
  
  \bibitem{wang} 
  Q.~Wang, C.~Hanhart and Q.~Zhao,
  Phys.\ Rev.\ Lett.\  {\bf 111}, 132003 (2013).
  
  \bibitem{carlos} 
  C.~Hidalgo-Duque, J.~Nieves, A.~Ozpineci and V.~Zamiralov,
  Phys.\ Lett.\ B {\bf 727}, 432 (2013).
  
  \bibitem{mknno1} 
  A.~Martinez Torres, K.~P.~Khemchandani, F.~S.~Navarra, M.~Nielsen and E.~Oset,
  Phys.\ Rev.\ D {\bf 89}, 014025 (2014)
  
   \bibitem{liang} 
  W.~H.~Liang, J.~J.~Xie, E.~Oset, R.~Molina and M.~Dšring,
  Eur.\ Phys.\ J.\ A {\bf 51}, 58 (2015).
 
  \bibitem{guo} 
  F.~K.~Guo, C.~Hanhart, Q.~Wang and Q.~Zhao,
  Phys.\ Rev.\ D {\bf 91}, 051504 (2015).
  
  \bibitem{albuquerque1} 
  R.~M.~Albuquerque, M.~E.~Bracco and M.~Nielsen,
  Phys.\ Lett.\ B {\bf 678}, 186 (2009).

  \bibitem{albuquerque2} 
  R.~M.~Albuquerque, J.~M.~Dias, M.~Nielsen and C.~M.~Zanetti,
  Phys.\ Rev.\ D {\bf 89}, 076007 (2014).

\bibitem{Zhigangwang}
  Z.~G.~Wang,
  Eur.\ Phys.\ J.\ C {\bf 74} (2014) no.7,  2963;
  Z.~G.~Wang,
  Eur.\ Phys.\ J.\ C {\bf 63} (2009) 115;
  Z.~g.~Wang and Y.~f.~Tian,
  Int.\ J.\ Mod.\ Phys.\ A {\bf 30} (2015) 1550004.
   
  \bibitem{kleiv} 
  R.~T.~Kleiv, T.~G.~Steele, A.~Zhang and I.~Blokland,
  Phys.\ Rev.\ D {\bf 87}, 125018 (2013).
  
  \bibitem{mknno2} 
  A.~Martinez Torres, K.~P.~Khemchandani, M.~Nielsen, F.~S.~Navarra and E.~Oset,
  Phys.\ Rev.\ D {\bf 88}, 074033 (2013).
  
  \bibitem{mknn2} 
  K.~P.~Khemchandani, A.~Martinez Torres, M.~Nielsen and F.~S.~Navarra,
  Phys.\ Rev.\ D {\bf 89}, 014029 (2014) 
  
  \bibitem{mo} 
  Z.~Mo, C.~Y.~Cui, Y.~L.~Liu and M.~Q.~Huang,
  Commun.\ Theor.\ Phys.\  {\bf 61}, 501 (2014).
  
   
  \bibitem{belle}
 P. Pakhlov {\it et al}. (Belle Collaboration), Phys. Rev. Lett.
{\bf 100}, 202001 (2008).

  \bibitem{raquel} 
  R.~Molina and E.~Oset,
  Phys.\ Rev.\ D {\bf 80}, 114013 (2009).
  
  \bibitem{li} 
  B.~Q.~Li and K.~T.~Chao,
  Phys.\ Rev.\ D {\bf 79}, 094004 (2009).
  
   \bibitem{yang} 
  Y.~c.~Yang, Z.~Xia and J.~Ping,
  Phys.\ Rev.\ D {\bf 81}, 094003 (2010).
  
  \bibitem{cao} 
  L.~Cao, Y.~C.~Yang and H.~Chen,
  Few Body Syst.\  {\bf 53}, 327 (2012).
  
  \bibitem{zhu} 
  R.~Zhu,
  Phys.\ Rev.\ D {\bf 92}, 074017 (2015).
  
 
 \bibitem{cdf}
 T. Aaltonen {\it et al}. (CDF Collaboration), Phys. Rev. Lett. {\bf 102}, 242002 (2009)
  
  \bibitem{cms}
  S.~Chatrchyan {\it et al.} [CMS Collaboration],
  Phys.\ Lett.\ B {\bf 734}, 261 (2014).
  
\bibitem{d0} 
  V.~M.~Abazov {\it et al.} [D0 Collaboration],
  Phys.\ Rev.\ Lett.\  {\bf 115}, no. 23, 232001 (2015).
  
  \bibitem{lhcb}
  R.~Aaij {\it et al.} [LHCb Collaboration],
  Phys.\ Rev.\ Lett.\  {\bf 118} (2017) no.2,  022003;   R.~Aaij {\it et al.} [LHCb Collaboration],
  Phys.\ Rev.\ D {\bf 95} (2017) no.1,  012002
  
  
   \bibitem{eef3} 
  E.~van Beveren and G.~Rupp,
  arXiv:0906.2278 [hep-ph].
  
  \bibitem{liuoka} 
  X.~H.~Liu and M.~Oka,
  Phys.\ Rev.\ D {\bf 93}, 054032 (2016).

  \bibitem{gonzalez} 
  P.~Gonz\'alez,
  Phys.\ Rev.\ D {\bf 92}, 014017 (2015).
  
  \bibitem{liuzhu} 
  X.~Liu and S.~L.~Zhu,
  Phys.\ Rev.\ D {\bf 80}, 017502 (2009)
  Erratum: [Phys.\ Rev.\ D {\bf 85}, 019902 (2012)].
  
    
  \bibitem{lebed} 
  R.~F.~Lebed and A.~D.~Polosa,
  arXiv:1602.08421 [hep-ph].


  
    \bibitem{HidalgoDuque:2012pq} 
  C.~Hidalgo-Duque, J.~Nieves and M.~P.~Valderrama,
  Phys.\ Rev.\ D {\bf 87}, 076006 (2013).


  
   \bibitem{svz}
M.A. Shifman, A.I. and Vainshtein and V.I. Zakharov,
Nucl. Phys. B {\bf 147}, 385 (1979);

\bibitem{reviews1}
 P.~Colangelo and A.~Khodjamirian,
  In *Shifman, M. (ed.): At the frontier of particle physics, vol. 3* 1495-1576
  [hep-ph/0010175].

 \bibitem{reviews2}
S. Narison, {\it QCD as a theory of hadrons,
Cambridge Monogr. Part. Phys. Nucl. Phys. Cosmol.} {\bf 17}, 1 (2002); {\it QCD spectral sum rules ,  World Sci. Lect. Notes Phys.} 
{\bf 26}, 1 (1989);
{ Acta Phys. Pol.} B {\bf 26}, 687 (1995); { Riv. Nuov. Cim.} {\bf 10N2}, 1
(1987); { Phys. Rept.} {\bf 84}, 263 (1982).

 \bibitem{reviews3} S.~Narison, Phys.\ Lett.\ B {\bf 216}, 191 (1989); {\bf 341}, 73 
(1994); {\bf 361}, 121 (1995), {\bf 387}, 162 (1996); {\bf 466}, 345 (1999); 
{\bf 624}, 223 (2005).

\bibitem{Zanetti:2016wjn} 
  C.~M.~Zanetti, M.~Nielsen and K.~P.~Khemchandani,
  Phys.\ Rev.\ D {\bf 93}, 096011 (2016).
 
 \bibitem{pascual}
 P.~Pascual, R. Tarrach, {\it QCD: Renormalization for the practitioner}, Springer-Verlag Berlin (1984).

\bibitem{Bondar:2011ev}
  A.~E.~Bondar, A.~Garmash, A.~I.~Milstein, R.~Mizuk and M.~B.~Voloshin,
  Phys.\ Rev.\ D {\bf 84} (2011) 054010.


\bibitem{Voloshin:2011qa}
  M.~B.~Voloshin,
  Phys.\ Rev.\ D {\bf 84} (2011) 031502.


\bibitem{Reinders:1984sr}
  L.~J.~Reinders, H.~Rubinstein and S.~Yazaki,
  Phys.\ Rept.\  {\bf 127} (1985) 1.
  
 
 
  \bibitem{ioffe}
  B.~L.~Ioffe and A.~V.~Smilga,
  Nucl.\ Phys.\ B {\bf 232} (1984) 109.
  doi:10.1016/0550-3213(84)90364-X


\bibitem{eidemuller}
M.~Eidemuller, F.~S. Navarra, M.~Nielsen and R.~R. da Silva, 
Phys.\ Rev.\ D {\bf 72}, 034003 (2005).

 \bibitem{Bracco:2011pg} 
  M.~E.~Bracco, M.~Chiapparini, F.~S.~Navarra and M.~Nielsen,
  Prog.\ Part.\ Nucl.\ Phys.\  {\bf 67}, 1019 (2012).
 



\bibitem{Aliev2010} 
  T.~M.~Aliev, K.~Azizi and M.~Savci,
  Phys.\ Lett.\ B {\bf 690}, 164 (2010).

  \bibitem{Alhendi2015} 
  H.~A.~Alhendi, T.~M.~Aliev and M.~Savcõ,
  JHEP {\bf 1604}, 050 (2016).

\bibitem{Li2015} 
  Z.~Y.~Li, Z.~G.~Wang and G.~L.~Yu,
  Mod.\ Phys.\ Lett.\ A {\bf 31}, no. 06, 1650036 (2016).
  
\bibitem{Giacosa:2005bw}
  F.~Giacosa, T.~Gutsche, V.~E.~Lyubovitskij and A.~Faessler,
  Phys.\ Rev.\ D {\bf 72} (2005) 114021.
  

\bibitem{oller} 
  J.~A.~Oller and E.~Oset,
  Nucl.\ Phys.\ A {\bf 620}, 438 (1997)
  Erratum: [Nucl.\ Phys.\ A {\bf 652}, 407 (1999)]

  \end{thebibliography}
\end{document}